\documentclass[5p]{elsarticle}

\usepackage{hyperref}
\usepackage{dblfloatfix} 
\usepackage{float}
\usepackage{graphicx}
\usepackage{soul} 
\usepackage{xcolor} 
\usepackage{caption}
\usepackage{bm}
\usepackage{svg}
\usepackage{amsmath}
\usepackage[normalem]{ulem}
\usepackage{upgreek}
\usepackage{makecell}
\usepackage{multirow}

\usepackage{nicefrac}

\journal{Surface and Coatings Technology}

\begin{document}


\sloppy

\begin{frontmatter}

\title{Strain rate sensitivity of a Cu/Al$_2$O$_3$ multi-layered thin film}


\author[myaddress1,myaddress3]{Szilvia Kal\'{a}cska\corref{mycorrespondingauthor2}}\cortext[mycorrespondingauthor2]{Corresponding author} \ead{szilvia.kalacska@cnrs.fr}
\author[myaddress3]{Laszlo Pethö}
\author[myaddress1]{Guillaume Kermouche}
\author[myaddress3]{Johann Michler}
\author[myaddress2]{Péter Dusán Ispánovity}

\address[myaddress1]{Mines Saint-Etienne, Univ Lyon, CNRS, UMR 5307 LGF, Centre SMS, 158 cours Fauriel 42023 Saint-Étienne, France}
\address[myaddress3]{Empa, Swiss Federal Laboratories for Materials Science and Technology, Laboratory for Mechanics of Materials and Nanostructures, Feuerwerkerstrasse 39, 3602 Thun, Switzerland}
\address[myaddress2]{ELTE Eötvös Loránd University, Department of Materials Physics, Pázmány Péter sétány 1/a, 1117 Budapest, Hungary}

\begin{abstract}
To study the size and strain rate dependency of copper polycrystalline microstructures, a multi-layered copper/Al$_2$O$_3$  thin film was deposited on a Si substrate using a hybrid deposition system (combining physical vapour and atomic layer deposition). High temperature treatment was applied on the ``As Deposited" material with ultrafine-grained structure to increase the average grain size, resulting in a ``Heat Treated" state with microcrystalline structure. Focused ion beam milling was employed to create square shaped micropillars with two different sizes, that were subjected to compressive loading at various (0.001/s -- 1000/s) strain rates. Differences in the strain rate sensitivity behavior manifesting at low and high strain rates are discussed in the context of the pillar diameters and the grain size of the deformed samples. The Al$_2$O$_3$ interlayer studied by transmission electron microscopy showed excellent thermal stability and grain boundary pinning by precipitation, also resulting in the homogeneous deformation of the pillars and preventing shear localization. Geometrically necessary dislocation densities estimated by high (angular) resolution electron backscatter diffraction presented inhomogeneous dislocation distribution within the deformed pillar volumes, that is attributed to the proximity of the sample edges. Finally, the Al$_2$O$_3$ interlayers successfully suppressed any possible recrystallization processes, contributing to the excellent film stability, that makes the proposed coating ideal to be operating under extreme conditions.

\end{abstract}

\begin{keyword}
thin films, micromechanical testing, microstructure, high strain rate, plastic deformation, size effect
\end{keyword}

\end{frontmatter}

\section{Introduction}

Thin films are employed by a large variety of applications. Protective coatings can improve corrosion or scratch resistance, they can hold decorative purposes (to change the color/roughness of the surface), or they can have other functional objectives. Creating multi-layered thin films with alternating dissimilar sublayers promises unusual (electrical \cite{Wu.2017}, thermal \cite{Cortes.2023}, optical \cite{Wang.2018}, etc.) properties to be experimentally investigated. Such systems, where grain size and texture can be controlled by the deposition / annealing process, represent an outstanding opportunity to focus on a few key aspects of the deformation processes driven by the collective behavior of dislocations.

Copper coating is important in several applications due to its excellent conductivity, corrosion resistance \cite{Yu.2007}, and antimicrobial properties. Copper is the most widespread choice as a conducting material for connections in integrated circuits with submicron features \cite{Giroire.2017}. The microstructure of the coating not only influences the thermal and electrical conductivity \cite{Mech.2011}, but also its mechanical properties, too. Therefore, it is important to study the deposited coating from the structural engineering point of view too.

Hybrid thin film coatings emerged due to their enhanced mechanical properties, such as in Al/Al$_2$O$_3$ depositions \cite{xie.2020, edwards.2022}. Here, the combination of physical vapor deposition (PVD) is combined with atomic layer deposition (ALD). PVD is employed to grow high purity thick layers in a low oxygen environment, while ALD is used to obtain precise thickness control over thinner (typically in the nm range) layers.

Investigating the mechanical stability under various extreme conditions is crucial for defining the margins of the coatings' applicability. Recent innovations in experimental micromechanical testing now enables probing at a wide range of deformation speeds \cite{nadia.2020, raj.2021, ventura.2022, schwiedrzik.2022, ventura.2023} at the relevant (application) scale. The two essential parameters for understanding deformation kinetics are the strain rate sensitivity (SRS, $m$) of the flow stress and the activation volume ($\Omega$) \cite{Cheng.2005}. Analyzing these two parameters can provide a clearer understanding of the rate-controlling deformation mechanisms. However, there is limited experimental information on the high strain rate (HSR) regime mechanical behavior as a function of size \cite{Soni.2020, Langdon.2022} and microstructure in the literature \cite{Zhang.2014, kalacska.2022}.

\begin{figure*}[!hb]
    \centering
    \includegraphics[width=0.95\textwidth]{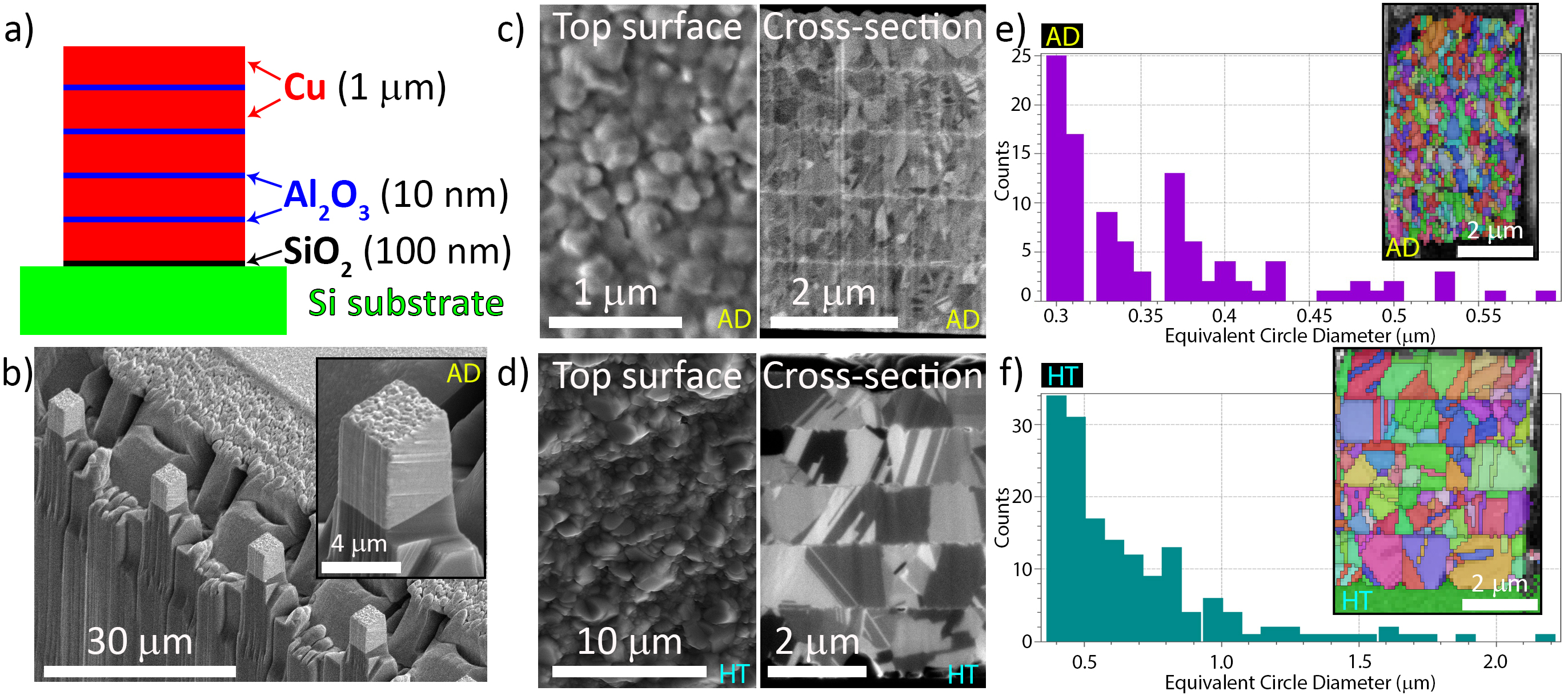}
    \caption{\textbf{a)} Schematics of the deposited multi-layered thin film coating, indicating 5 layers of Cu (PVD), separated by thin Al$_2$O$_3$ (ALD) layers. \textbf{b)} Overview of the FIB-milled batch of the ``As Deposited - AD" micropillars. The inset shows one pillar with higher magnification, where the brighter volume corresponds to the deposited thin film. \textbf{c)} Top surface and cross-sectional images of the AD coating. \textbf{d)} Top surface and cross-sectional images of the ``Heat Treated - HT" coating. \textbf{e)} Grain size distribution of the AD sample, based on cross-sectional EBSD measurements (inset). \textbf{f)} Grain size distribution of the HT sample, based on cross-sectional EBSD measurements (inset).
    \label{fig:01}}
\end{figure*}

The aim of this study is, therefore, to design a hybrid (Cu/Al$_2$O$_3$) thin film and tailor its microstructure by heat treatment to study the SRS behavior as a function of sample size and microstructure. The applied material can be considered as “model material”, since the aim of this work was to provide a straightforward method for studying certain high strain rate properties (e.g. strain rate sensitivity of the yield stress as a function of microstructure), due to Cu’s well-understood characteristics. Since the preparation methodology can be consistently repeated, the fundamental properties and underlying mechanisms can be representative to a broader class of thin films, making it a useful proxy for understanding similar systems. The purpose of the Al$_2$O$_3$ interlayer was to limit grain growth between the subsequent Cu depositions, allowing to investigate the ``As Deposited" (ultrafine-grained) coating parallel to its ``Heat Treated" counterpart containing
microcrystalline grains.

\section{Materials \& Methods}

\begin{figure*}[!hb]
    \centering
    \includegraphics[width=0.95\textwidth]{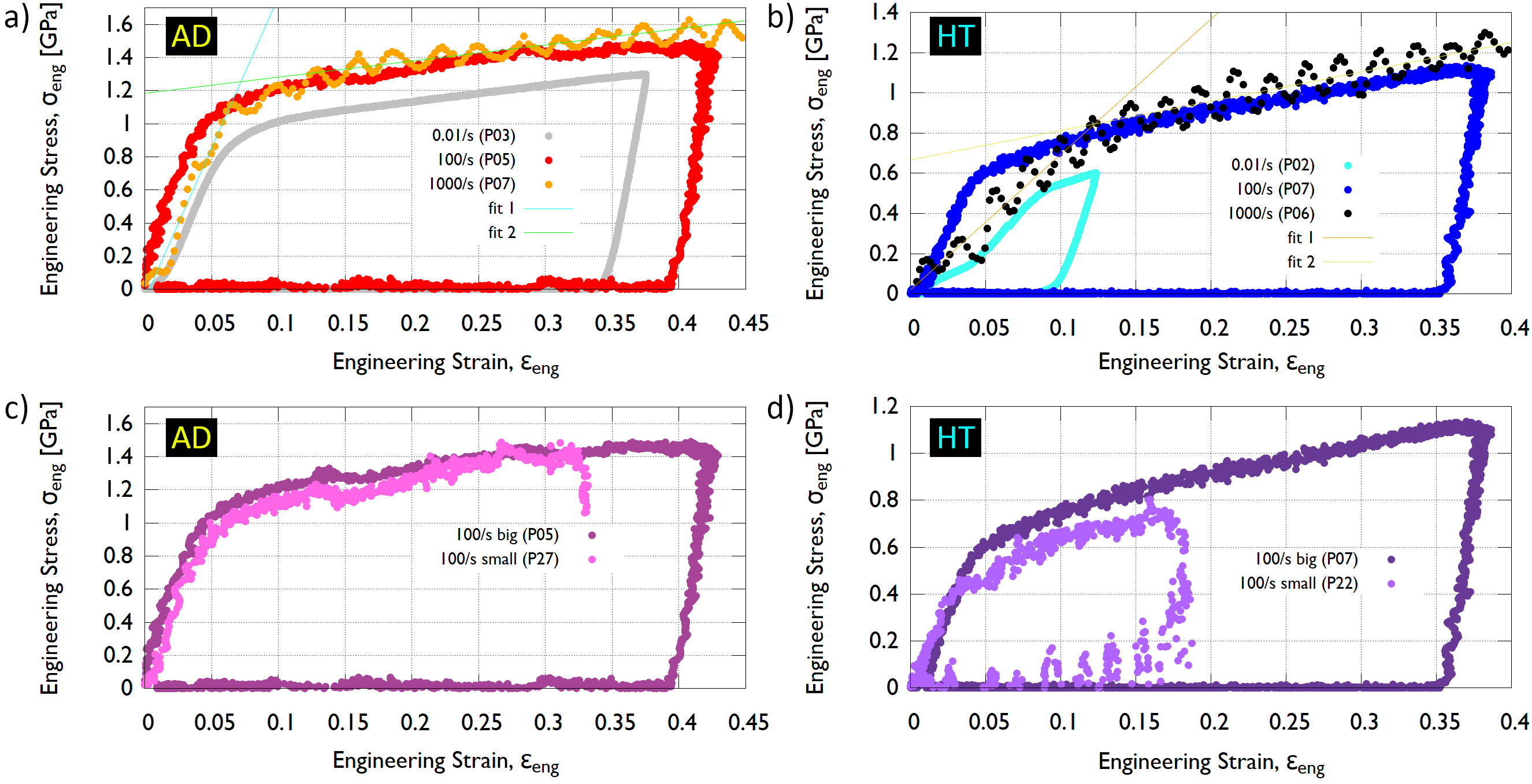}
    \caption{\textbf{$\sigma_{eng}$ - $\varepsilon_{eng}$ curves at various strain rates.} \textbf{a)} AD pillars, shown with the linear fits for 1000/s curve (pillar P07) for the yield stress determination. \textbf{b)} HT pillars, shown with the linear fits for 1000/s curve (pillar P06) for the yield stress determination. \textbf{c)} AD pillars with two different sizes at SR=100/s. \textbf{d)} HT pillars with two different sizes at SR=100/s.}
    \label{fig:02}
\end{figure*}

\subsection{Sample preparation}
For the creation of the desired multilayers, a hybrid thin film deposition system combining both ALD and PVD methods was applied (Swiss Cluster). The sequential deposition consisted of alternating (5 times) $\sim 1$ $\upmu$m thick pure Cu separated by 10 nm thin Al$_2$O$_3$ interlayers, deposited on a Si (100) substrate coated with a 100 nm thick SiO$_2$ diffusion barrier (Fig. \ref{fig:01}a) at room temperature. Copper was deposited by DC magnetron sputtering in current control mode. Two magnetrons were used, identical by design, mounted 180$^{\circ}$ apart within the chamber. The current on both was set to 200 mA, which resulted in 350 V of plasma potential and an average of 70 W plasma power. The base pressure prior deposition was better than $5\times 10^{-7}$ mbar. Targets with 50 mm diameter and 3 mm thickness were sputtered, supplied by HMW-Hauner GmbH, with a purity of 99.99\%. The substrate was kept at room temperature. During deposition, 15 sccm argon was released into the chamber, and the process pressure was held at $5\times 10^{-3}$ mbar by a throttle valve. It is important to note that the key advantage of the Swiss Cluster equipment is that the vacuum was maintained between PVD and ALD transitions, preventing the formation of a native oxide layer. Furthermore, these layers were prepared using the classical ALD method by alternating trimethylaluminum (TMA) and water as precursors \cite{Puurunen.2005}. This methodology was employed deliberately as a baseline for comparison and validation, adhering to widely accepted practices in the field \cite{Cremers.2019, Oviroh.2019}. Cross-sectional electron backscatter diffraction (EBSD) confirmed the average grain size of this ``As Deposited - AD" film to be $\overline{d^{AD}}\sim 120$ nm (area weighted mean, see in Fig. \ref{fig:01}c,e), qualifying this coating as ultrafine-grained (UFG).

Heat treatment was performed in a furnace at 800$^{\circ}$C for 4 hours in Ar gas to evade oxidation, that resulted in the ``Heat Treated - HT" sample. The roughness of the film surface along with the average grain size ($\overline{d^{HT}}\sim 1.1$ $\upmu$m) has increased considerably, as it can be seen in Fig.~\ref{fig:01}d,f. The increased grain size defines this coating having microcrystalline (MC) structure.

In order to investigate the cross-section of the coating a sharp edge was prepared by broad beam Ar ion polishing (Jeol IB-1953CP) with beams of 6 kV (60 mins) and 2 kV (60 mins). For the micromechanical experiments square based taper-free micro-samples were created close to the sharp edge by focused ion beam (FIB) using a FEI Helios 600i and a Tescan Lyra3 Ga$^+$ scanning electron microscope (SEM) system (Fig.~\ref{fig:01}b). Sequential preparatory beams of 30 kV 21 nA (rough milling for removing large amount of material around the region of interest) down to $2.5-0.4$ nA polishing were utilized with a 45$^{\circ}$ incident angle direction with respect to the prepared surfaces. Two batch of different sized cuboid pillars ($\sim 4.2$ $\upmu$m and $\sim 2.9$ $\upmu$m side length) were fabricated to study possible size related effects in the SRS.

\subsection{Micromechanical testing}

Micro-samples were deformed using an Alemnis ASA nanodeformation system equipped with three different load cells (strain gage based standard and mini load cells for quasi-static strain rates (SR), piezo-based SmarTip for high SRs). For pillar compressions, conductive diamond flat punches (Synton AG) with diameters of 5 $\upmu$m and 10 $\upmu$m were utilized. The recorded load-displacement curves were baseline (drift) and compliance corrected using the AMMDA evaluation software. During HSR testing, due to the high acceleration following the tip actuation (achieving up to 6 mm/s tip velocity), an oscillating noise appeared, superimposed to the signal (``ringing effect"). In order to determine the yield stresses $\sigma_\mathrm y$ in such difficult cases, the elastic and plastic parts of the stress-strain curves were fitted using a linear equation, and their intersection was used to obtain the apparent yield stress (see Fig.~\ref{fig:02}a,b and Suppl.~Fig.~S\ref{fig:S01} for the full process). This method enabled to avoid further data filtering or averaging in the $\sigma_\mathrm y$ determination. For the consistency of the yield stress values at low and high strain rates, all mechanical data were treated the same way. Additionally, the more common 1\% plastic strain offset method was also employed to assess the SRS behaviour, that is provided in Suppl. Fig. S\ref{fig:s08} for comparison.

\subsection{Microstructure characterization}
EBSD was employed to get statistical information on the grains of the samples. A Zeiss Supra 55 VP FEG-SEM equipped with an Oxford Instruments Symmetry 2 detector was used to record the EBSD diffraction patterns (2x2 binning, frame averaging: 4, step: 80-100 nm). High (angular) resolution (HR-) EBSD was utilized for the geometrically necessary dislocation (GND) density estimation prior to and following the deformation. This technique is based on image cross-correlation, linking each point on the map to a reference stress state \cite{Wilkinson.2006}. The stored patterns were evaluated using the BLG Vantage CrossCourt Rapide v4.6 HR-EBSD software. \textit{Post mortem} FIB-assisted 3D tomography \cite{kalacska2020_3d} was performed on two pillars (HT big pillars, deformed at 0.01/s and 1000/s) to study the effect of SR on the microstructure close to the surface and inside the deformed volume. 

Transmission electron microscopy (TEM) was used to gain information on the stability of the hybrid layers after high temperature processing. As a result of the heat treatment, the grains were expected to grow, that can disturb the continuity of the deposited thin (10 nm) oxide layer. For this, a Jeol-ARM200F Cold FEG NeoARM Cs-corrected TEM was utilized at 200 kV. This instrument is equipped with an energy dispersive spectroscopy (EDS) detector (SDD CENTURIO-X) too. Multiple TEM-lamellae were prepared by FIB milling, following the standard preparation recipe, using sequential settings from 30 kV, 21 nA--40 pA (rough milling and liftout) down to 5 kV--1 kV, 20 pA (thinning and final cleaning).

\begin{figure*}[!hb]
    \centering
    \includegraphics[width=0.95\textwidth]{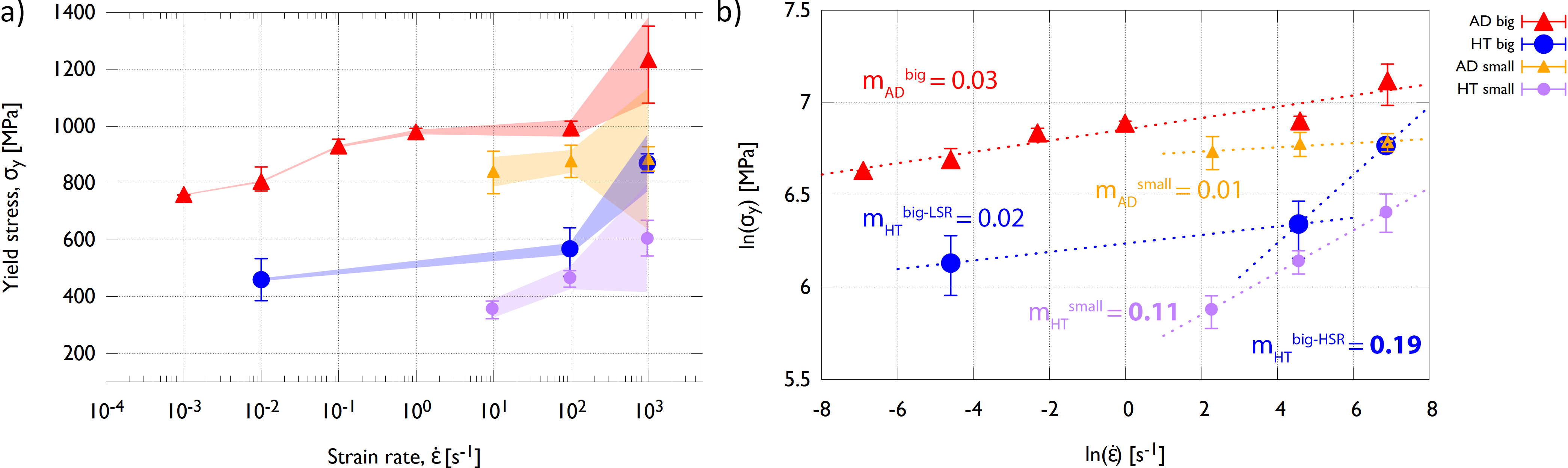}
    \caption{\textbf{Strain rate sensitivity of the tested pillars.} \textbf{a)} Strain rate (log. scale) -- yield stress (linear scale), and \textbf{b)} the corresponding logarithmic representation for the SRS-factor ($m$) determination (based on Eq.\ref{eq:srs}). Error bars are an indication of the spreading of the calculated yield stresses from repeatability tests. The shaded area shows the amount of noise at each strain rate step, while the $m$ values are determined by the fitted dotted lines.
    \label{fig:03}}
\end{figure*}

\section{Results}

The engineering stress ($\sigma_\mathrm{eng}$) - engineering strain ($\varepsilon_\mathrm{eng}$) curves resulting from several pillar compressions at various strain rates ($0.001\mathrm{/s} \leq \dot{\varepsilon} \leq 1000\mathrm{/s}$) are shown in Fig.~\ref{fig:02}. It is expected that the AD pillars exhibit higher yielding than the HT specimen, according to grain boundary strengthening. This phenomenon is confirmed in Fig. \ref{fig:02}a,b for the two types of coatings (AD and HT). Work hardening seems not to be affected by the SR change. It is important to note that in the HT samples, due to the increased coating roughness, the uneven surface can introduce an artefact to the mechanical data. As it can be observed in Fig.~\ref{fig:02}b (P02, $\dot{\varepsilon}=0.01$/s), if the surface of the pillar contains a part exceeding the average height, upon contact with the flat punch, this grain will start to deform, resulting in a less steep elastic loading (due to the fact that the $\sigma_\mathrm{eng} - \varepsilon_\mathrm{eng}$ curves are calculated with the full pillar dimensions). However, once the surface is levelled by the indenter tip, the elastic loading part is recovered while the full pillar cross-section is being compressed beyond this point. The primarily deformed grain might locally modify the deformation state of the neighboring grains, however, due to the multilayered nature of the thin film, we expect to see negligible influence of this effect in the global mechanical behavior. To avoid this, either \emph{i)} additional FIB polishing could be applied that levels the pillar surface, or \emph{ii)} a Pt deposition cap may be added. Both solution could potentially modify the recorded mechanical data \cite{edwards.2022}, so in this work, we avoid using any of the aforementioned corrections, and restrict ourselves to the analysis of the unmodified HT coating.

Smaller pillars exhibit lower yielding, that can be attributed to the \emph{i)} surface damage caused by FIB milling \cite{xiao.2017, xiao.2019} or \emph{ii)} it can be a grain size effect enhancing strain localization in the smaller HT specimen. As it was mentioned previously, to evade tapering, a higher incident angle of the bombarding ions were utilized, that may result in a higher concentration of implanted Ga close to the pillar surfaces. In smaller pillars, the effect of Ga presence is enhanced due to the increased surface-to-volume ratio. This explanation is the most probable cause of the slight yield stress decrease in smaller pillars rather than other  microstructure related effects, since it is apparent in both the AD (UFG) and HT (MC) pillars in a similar manner at both low and high strain rates (see in Fig. \ref{fig:03}).

In order to determine the SRS-factor ($m$), the following equation is used:

\begin{equation}\label{eq:srs}
    m=\frac{\partial (\ln{\sigma_\mathrm y})}{\partial (\ln{\dot{\varepsilon}})}.
\end{equation}

In Fig. \ref{fig:03}a, the determined average $\sigma_\mathrm y$ values are plotted on a linear scale as a function of $\dot{\varepsilon}$ on a logarithmic scale. The error bars show the spreading of the calculated values from repeated pillar compression tests (typically three), that presents good repeatability of the experiments at almost all applied strain rates. As a result of the employed linear fitting method, $\sigma_\mathrm y$ can be determined fairly consistently.  The shaded area represents the noise during loading, indicating that when a load cell is switched from strain gage based ($\dot{\varepsilon} \leq $ 1/s) to piezoelectric based detection ($\dot{\varepsilon} \geq $ 10/s), the noise increases significantly. The drawback of using a noisier (but much stiffer) load cell \cite{raj.2021} is compensated by the ability to go beyond the previously achievable deformation speeds and sampling rates. The noise becomes extremely large then the load signal starts to have a overwhelming oscillation due to the ringing effect ($\dot{\varepsilon} = $ 1000/s). To reduce this outcome originating from the immense acceleration during the operation (reaching 6 mm/s tip velocity upon impact), certain pillars were compressed in a way that the tip acceleration was initiated far ($\sim 14$ $\upmu$m) away from the surface, anticipating that at this distance the tip resonance decreases to minimal upon impact, when the flat punch is already travelling at its maximum speed \cite{Breumier.2020}. To further improve the mechanical data, some pillars were compressed to very high strains to allow the tip to drive through these pillars at a constant speed without stopping, thereby completely destroying the samples and preventing them from being examined by \emph{post mortem} EBSD analysis (hence the notation ``pancake" in Suppl.~Fig.~S\ref{fig:S04}-S\ref{fig:S06}).

Strain rate sensitivity is plotted in Fig.~\ref{fig:03}b for all the tested pillars. The fitted dotted lines are used to determine the $m$ values, calculated using Eq.~\ref{eq:srs}. At lower strain rates, the HT samples are expected to behave similarly as the AD samples based on the work of Ramachandramoorthy and Kalacska et al. (for more details see the \emph{Discussion} section.) \cite{kalacska.2022}, therefore, only one additional $\dot{\varepsilon}$ state was tested in this regime. Increased repeatability error in the HT samples are attributed to the surface roughness increase, as stated previously.

\begin{figure*}[!ht]
    \centering
    \includegraphics[width=0.95\textwidth]{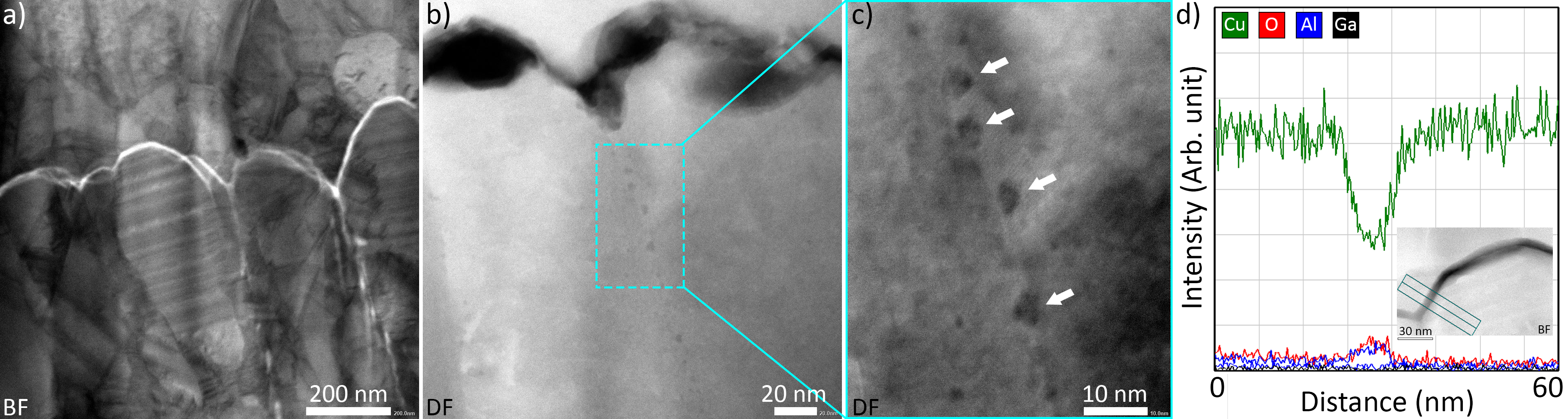}
    \caption{\textbf{TEM results.} \textbf{a)} Bright field (BF) image of a lamella lifted out from the HT sample, showing the Cu grain structure, together with the Al$_2$O$_3$ interlayer (white contrast). \textbf{b)} Dark field (DF) image of an area between two larger grains, here the Al$_2$O$_3$ layer appears to be black. \textbf{c)} Enlarged section of the image in b), showing inter-granular Al$_2$O$_3$ precipitates (white arrows). \textbf{d)} Chemical composition analysis by EDS performed on the area in the inset measures the Al$_2$O$_3$ layer to be $\sim 13.5$ nm thick.}
    \label{fig:04}
\end{figure*}

In order to understand the effect of heat treatment on the Cu/Al$_2$O$_3$ multi-system, TEM-lamellae were prepared from the HT sample via FIB milling. The results shown in Fig.~\ref{fig:04} confirm the enlarged grains due to high temperature ageing. Precipitates along grain boundaries were observed with sizes varying between 2-5 nm (Fig.~\ref{fig:04}c). EDS line profile collected through a Cu/Al$_2$O$_3$/Cu interface confirmed the integrity of the Al$_2$O$_3$ deposition after heat treatment, and the estimated layer thickness was calculated to be $\sim 13.5$ nm (Fig.~\ref{fig:04}d). It is mentioned that this result may depend on the thickness of the lamella.

\begin{figure}[!ht]
    \centering
    \includegraphics[width=0.45\textwidth]{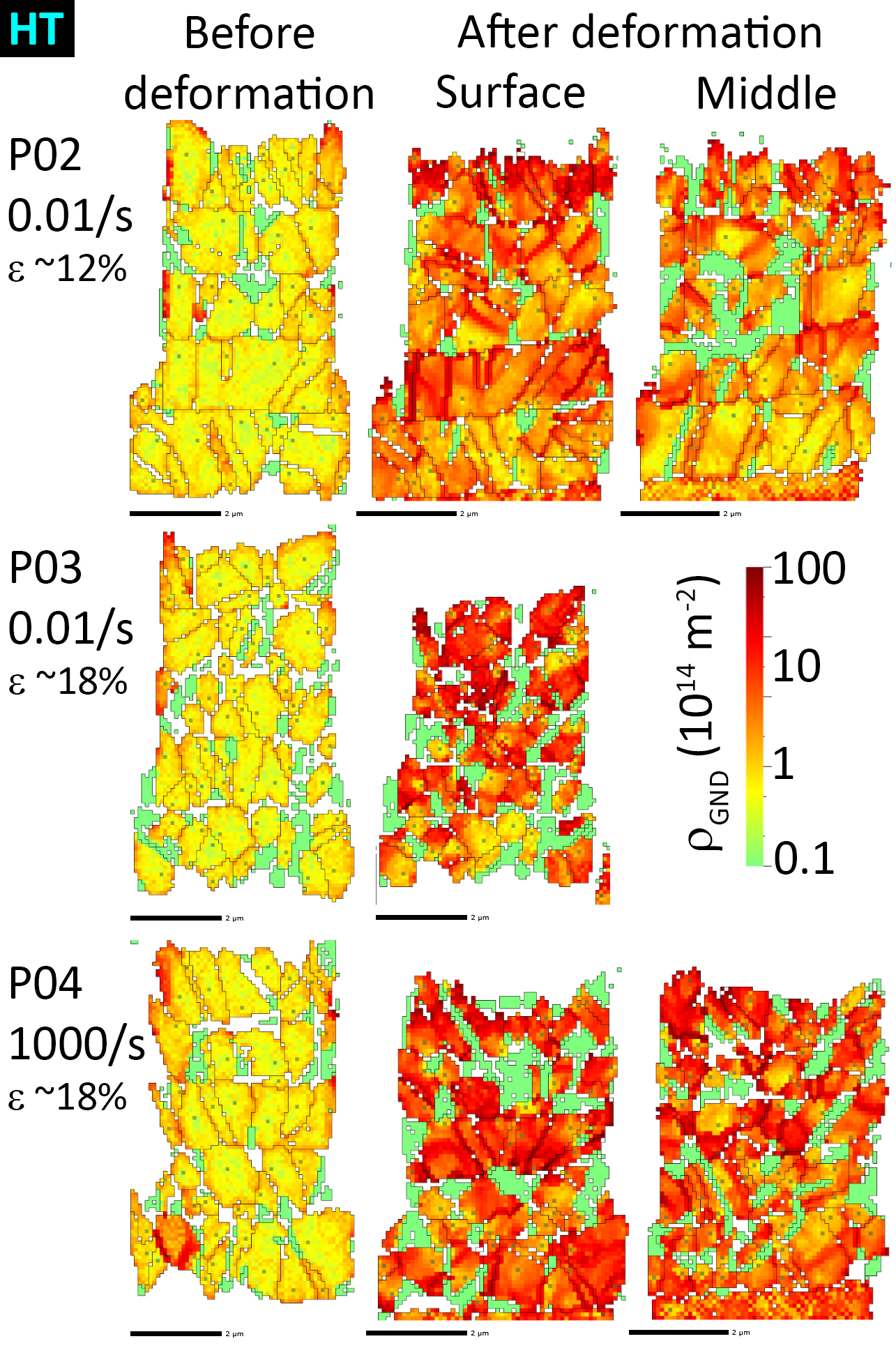}
    \caption{\textbf{HR-EBSD results.} Maps were measured on the surface before and after deformation ($\varepsilon$). Pillars P02 and P04 were subjected to serial cross-sectioning by FIB, where the middle of the pillars (about 1.6 $\upmu$m away from the first slice) were also imaged. Black scale bars corresponds to 2 $\upmu$m.}
    \label{fig:05}
\end{figure}

Finally, the effect of SR on the evolved dislocation structure was investigated by cross-sectional HR-EBSD. The pillars' surfaces were mapped before and after deformation, and the resulting GND density maps are plotted in Fig.~\ref{fig:05}. Here, the small grains (containing less than 20 pixels) were excluded from the analysis (green areas). Also, patterns with lower band contrast (BC) value than a given threshold (BC $<100$) were also removed from the evaluation (i.e. where patterns overlap close to grain boundaries). Afterwards, each identified grain was assigned to a reference pattern, chosen from the same grain with the lowest kernel average misorientation value (black dots). GND densities ($\rho_\mathrm{GND}$) were then estimated based on 20 regions of interest for three pillars (P02 -- P04).

\section{Discussion}

Initially, the FIB-milled pillars were chosen to be square shaped for two main reasons: \emph{i)} to allow the monitoring of the $\rho_\mathrm{GND}$ evolution by HR-EBSD mapping, and \emph{ii)} to create perfect uniaxial loading conditions and identical cross-sections along the height of the pillars for more accurate mechanical data analysis. Looking at the already deformed samples (Suppl.~Fig.~S\ref{fig:S02}) it can be concluded that the pillars deform in a ``barrelling" manner, achieving the largest shape change closer to the top of the specimens at the low strain rate (LSR) regime (softer top). When a pillar is compressed at HSR ($\leq$ 100/s), the shape of the pillars become slightly different, as here the middle of the coating extrudes the most.

Slip plane localization is successfully suppressed by the consecutive Al$_2$O$_3$ layers in both the AD and HT samples, hence the resulting smooth $\sigma_\mathrm{eng}$ -- $\varepsilon_\mathrm{eng}$ curves in Fig.~\ref{fig:02}, that show the absence of large strain bursts (or stress drops). The surface analysis of the deformed pillars support these findings in Suppl.~Figs.~S\ref{fig:S02} and S\ref{fig:S03}. Flow stresses reach quite high average values ($\overline{\sigma_\mathrm y^\mathrm{AD}}=758$ MPa and $\overline{\sigma_\mathrm y^\mathrm{HT}}=460$ MPa) during quasi-static deformation, much higher than pure Cu in literature, approaching the regime of nanocrystalline copper \cite{Schiotz.2003, Guduru.2007}. This increased yielding could be partly attributed to the grain boundary (Hall-Petch) strengthening (using the already established $\overline{d}$ values for the AD and HT samples with the equation $\sigma_\mathrm{GB} = \sigma_0 +kd^{-0.5}$, with $\sigma_0^\mathrm{Cu} = 25$ MPa material constant and $k=0.12$ MPa m$^{1/2}$ strengthening coefficient \cite{smith2006foundations}, resulting in $\sigma_\mathrm{GB}^\mathrm{AD}=371$ MPa and $\sigma_\mathrm{GB}^\mathrm{HT}=139$ MPa theoretical values). Further strengthening mechanisms also play a key role in raising the $\sigma_\mathrm y$ values, such as dislocation and potentially (based on Fig.~\ref{fig:04}c) precipitation strengthening. As suggested by Fig.~\ref{fig:04}a, nanotwinning also occurs in this thin film system \cite{Sun.2018} that could be significant in the HSR behavior in case of possible dynamic recrystallization \cite{Tiamiyu.2022}. Additionally, high angle grain boundaries in the AD sample tend to run along the deformation axis of the pillars (somewhat columnar grain growth resulting in ellipsoid grain shapes), that will also increase the material's resistance towards compressive/tensile stresses.

When we compare the yield stress values to the 3D printed Cu UFG/MC system published in Ref. \cite{kalacska.2022}, we found these results to be in good agreement with both the AD/HT yielding, meaning that the multi-layered thin film proposed here is a fitting model material to further elaborate on the previously detected anomalous behavior of the polycrystalline Cu strain rate sensitivity. 

In Fig.~\ref{fig:03}, two trends in the SRS is discussed: \emph{i)} the monotonous $\sigma_\mathrm{y}$ increase with increasing SR in all studied samples at lower strain rates, and \emph{ii)} $\sigma_\mathrm{y}$ increase at 1000/s. The first behavior was already reported in the literature on similar materials (e.g. \cite{kalacska.2022}, where circular Cu polycrystalline pillars with 2 $\upmu$m gauge diameters were tested). There, it was found that pillars with UFG structure exhibited an anomalous SR saturation at $\dot \varepsilon \sim $ 500/s, modifying the SRS factor from $m_{\mathrm{UFG}}^{\mathrm{LSR}} =0.06$ to $m_{\mathrm{UFG}}^{\mathrm{HSR}} =0.003$. Here, we found similar behavior looking at the AD results between $0.1$ $s^{-1}$$< \dot{\varepsilon} \leq 100$ $s^{-1}$ (Fig. \ref{fig:03}a). However, $m$ remained constant in the smaller (2.9 $\upmu$m side length) pillars at HSR, leading to the conclusion that the pillar microstructure is not the only driving factor in the SRS behavior, but also the size of the sample has an additional influence too. In our AD thin films, the SRS factors were found to be close to the already reported values ($m_{\mathrm{AD}}^{\mathrm{LSR}} =0.03$, $m_{\mathrm{AD}}^{\mathrm{HSR}} =0.01$). When pillars were fabricated with bigger (4 $\upmu$m) side length, the SRS factor $m$ remained constant throughout the whole investigated $\dot{\varepsilon}$ regime.

Assuming that the active deformation mechanisms are temperature activated allows one to determine the activation volume characterizing the stress dependent part of the Gibbs potential. If deformation is governed by a single mechanism then the activation volume is expected to be constant at different applied stresses. The change in the activation volume, therefore, suggests a change in the dominant deformation mechanism. The apparent activation volume at the yield point is calculated by the following equation \cite{Schoeck.1965, Krausz.1976, gupta2020situ}:
\begin{equation}\label{eq:srs2}
    \Omega=\sqrt{3}k_\mathrm{B} T\frac{\partial(\ln{\dot{\varepsilon}})}{\partial (\ln{\sigma_\mathrm{y}})},
\end{equation}
where $k_\mathrm{B}$ and $T$ are the Boltzmann constant and the temperature, respectively. By using $b = 0.256$ nm as the Burgers vector of a \{111\}$<110>$ dislocation in copper, and $T=297$ K, the calculated activation volumes are $\Omega_{\mathrm{AD}}^{\mathrm{big}}\sim 15b^3$, $\Omega_{\mathrm{AD}}^{\mathrm{small}}\sim 49b^3$, and $\Omega_{\mathrm{HT}}^{\mathrm{small-HSR}}\sim 8b^3$. The measured volumes are in the same range as previous measurements on similar (UFG) materials \cite{Cheng.2005, Champion.2008}. Since in materials with complex microstructure, like the one studied here, usually several deformation mechanisms are active simultaneously, interpretation of activation volumes is rather challenging. If one assumes that there is a dominant mechanism, it 
suggests that in the LSR regime in the HT pillars, the deformation is controlled by the movement of dislocations ($1b^3-10b^3$) nucleated at high angle grain boundaries and at surface sources \cite{Jennings.2011}. In the small AD pillars, the increase in $\Omega$ suggests a collective dislocation dynamic process ($10b^3-100b^3$) \cite{Jennings.2011}, meaning that dislocations are created at multiple sources in large numbers upon impact, but their movement is restrained. To draw more detailed conclusions one would need to gain more information about the active deformation mechanisms either by, e.g., performing an \emph{in situ} TEM experiment \cite{gupta2020situ} or by analyzing complementary atomistic simulations \cite{zhang2022atomistic}. Other mechanisms could also play a significant role in the SRS behaviour (e.g. grain boundary diffusion-assisted processes influenced by creep-deformation \cite{Karanj.2012, Jonnal.2010}), that requires more targeted experimental studies to be performed.

Looking at Fig. \ref{fig:03}it can be seen that upon testing the pillars above 100/s, where the $\sigma_\mathrm{y}$ values unexpectedly increase in the HT samples, regardless of their size. Firstly, this apparent SRS change can originate from the increased noise levels during HSR testing. At LSRs, HT pillars exhibited similar SRS ($m_{\mathrm{HT}}^{\mathrm{LSR}} =0.02$) as the AD samples. This value matches the $m$ factor reported in Ref. \cite{kalacska.2022} for the MC structured pillars, where a monotonous SRS was also observed in this $\dot{\varepsilon}$ regime. However, upon testing the HT pillars at $\dot{\varepsilon}=1000$/s, the detected $m$ value has increased significantly ($m_{\mathrm{HT}}^{\mathrm{big-HSR}} =0.19$, $m_{\mathrm{HT}}^{\mathrm{small-HSR}} =0.11$), showing that at HSR, a transition in the dislocation dynamics may happen, resulting in a strain rate hardening regime. This crossover has already been proposed by simulations \cite{Fan.2021} and demonstrated by large scale impact testing \cite{Follansbee.1988, Kumar.2015}, but, to our knowledge, it is the first time to be presented at such small scale, during micromechanical testing. According to Ref. \cite{Kumar.2015}, at LSRs, dislocations are thermally activated, while when  $\dot{\varepsilon}$ exceeds $10^2$/s, dislocation velocities become so high that the phonon-drag (thermal elastic vibrations) on their motion can no longer be  neglected \cite{Meyers.1994}, making dislocations much more difficult to move, due to which we observe such an outstanding increase in the $\sigma_\mathrm{y}$ values. As the microstructure (and the initial dislocation density) plays a crucial part in the SRS, if such a mechanism is anticipated, this deviation will most likely present itself in the AD samples too, only at somewhat higher SRs than what is currently reachable by small scale experiments \cite{Mao.2018}. However, to have a definitive conclusion of the origin of this phenomenon, a more rigorous and systematic HSR testing is required.

\subsection{Microstructure evolution}

As it can be seen in Fig.~\ref{fig:04},and Suppl.~Figs.~S\ref{fig:S04} and S\ref{fig:S06}, the heat treatment has increased the initial grain size, that has caused the previously flat Al$_2$O$_3$ layers to become rippled, following the curvature of the grains growing in their vicinity. This process can also affect the thickness of the Al$_2$O$_3$ layer, too, as the increased grain surface can thin down the interlayer, just like the material of a balloon upon its inflation. One of the striking things to notice was that the grain growth has seemingly reached a final state and further grain growth was blocked even at longer ageing (72 hours was the longest achieved). Even after repeated attempts (initially at 600$^{\circ}$C, then 700$^{\circ}$C and finally 800$^{\circ}$C treatment using different testpieces), the grain growth process was halted. In this system, the Cu grain boundaries could not disappear due to an unknown phenomena. As a possible explanation, the diffusion of the Al (or Al$_2$O$_3$) has been proposed, by either \emph{i)} precipitation along grain boundaries, or \emph{ii)} atomic diffusion, as in an intra- or inter-granular manner. High resolution TEM imaging has confirmed the first hypothesis, showing that Al$_2$O$_3$ precipitates are located between large grains (marked by white arrows in Fig. \ref{fig:04}c), where the curvature of the oxide layer has significantly increased. In order to confirm the second hypothesis based on atomic level diffusion, further experiments will have to be performed, that is out of the scope of the current work. In summary, the aluminum-oxide layer has successfully served its purpose as a thickness (and grain size) controlling barrier, but if larger grains are needed to be designed in the future, the composition or the thickness of the interlayer will need to be modified in the currently investigated system. Regarding the role of the Cu/alumina interface, further high resolution studies are needed to understand the extent to which the crystalline-amorphous interface in a metal-ceramic hybrid system influences dislocation nucleation, mobility, and annihilation.

Furthermore, the decrease of the initial total dislocation density ($\rho_{\mathrm{Total}}$) is also anticipated as a result of high temperature processing. Consequently, hardening rates during deformation could be different for the AD and HT samples. However, it is difficult to separate the effect of $\rho_{\mathrm{Total}}$ difference from the influence of the grain boundaries (that also act as dislocation sources) on the yielding and hardening behaviors, therefore this aspect is not discussed further. On the other hand, the resulting dislocation densities from low and high strain rate compressive testing can be investigated in case of the HT samples, where grains are large enough to perform meaningful GND density analysis by HR-EBSD. As the average GND densities shown in Fig.~\ref{fig:05} and in Table \ref{tab:01}, the initial $\overline{\rho_{\mathrm{GND}}}=1.3 \times 10^{14}$ m$^{-2}$ substantially increases due to external loading. When we compare similar maximum strains at two different SRs ($\varepsilon \sim 18\%$, $\dot{\varepsilon} \in 0.01$/s, 1000/s), the resulting $\rho_{\mathrm{GND}}$ values are approximately 10 times higher at the surface. By performing \emph{post mortem} FIB cross sectioning, we confirmed that in the middle of these pillars ($\sim 1.6$ $\upmu$m below the surface), the average $\rho_{\mathrm{GND}}$ values are somewhat lower, that could be attributed to the ``surface effect", allowing facilitated dislocation nucleation (mainly due to the proximity of free surface and FIB damage) close to the facets of the pillars. This is in good agreement with previous GND density measurements on Cu micropillars \cite{kalacska.2020}, but it should be noted that during pillar compression,  only a fraction of the generated dislocations is geometrically necessary. Additionally, material extrusion happens at the edges without further material constraints, therefore more lattice rotation and more GND generation can occur.

\begin{table}[]
    \centering
    \begin{tabular}{|l|c|c|c|}\hline
     \multirow{2}{*}{Dataset} & Before def. & \multicolumn{2}{c}{After def. [m$^{-2}$]}\\
      & (Surface) [m$^{-2}$] & (Surface) & (Middle) \\\hline
     P02 & $1.2 \times 10^{14}$ & $8.0 \times 10^{14}$ & $5.4 \times 10^{14}$ \\    
     P03 & $1.0 \times 10^{14}$ & $1.1 \times 10^{15}$ &  -- \\    
     P04 & $1.7 \times 10^{14}$ & $1.3 \times 10^{15}$ & $1.1 \times 10^{15}$  \\\hline
     Average & $1.3 \times 10^{14}$ & -- & --  \\\hline
    \end{tabular}
    \caption{Average $\rho_\mathrm{GND}$ values based on the HR-EBSD results in Fig.~\ref{fig:05}.}
    \label{tab:01}
\end{table}

\subsection{Strain rate related microstructure transformation}

In Ref.~\cite{kalacska.2022}, when UFG pillars were compressed at HSR, some pillars exhibited grain coarsening in the gauge section, that was attributed to static recrystallization due to the extremely high dislocation densities, and not resulting from dynamical processes during the compression itself \cite{Andrade.1994, Thevamaran.2016}. In order to verify the appearance of recrystallization in this thin film (that could potentially modify the deposited coating's properties after impact), the deformed pillars were mapped by EBSD immediately ($\sim 2$ hours) after micromechanical testing (Suppl.~Figs.~S\ref{fig:S04}- S\ref{fig:S06}), and after ageing the samples under ambient conditions for 1.5 years (Fig. S\ref{fig:S07}). We found no evidence of abnormal grain growth in these pillars, regardless of the applied SR. The absence of recrystallization possibly originates from the applied Al$_2$O$_3$ interlayers, that conclusively not only exhibit excellent high temperature stability, but could potentially deactivate unwanted microstructure transformation, if the coating is to be operating under extreme conditions.

\section{Conclusions}

In summary, the created Cu/Al$_2$O$_3$ multi-layered thin film was successfully tested in its ``As Deposited" and ``Heat Treated" form, that allowed us to create a model material system to study strain rate sensitivity as a function of the micropillar size and microstructure. The 10 nm thin Al$_2$O$_3$ interlayers showed excellent thermal stability, even when long heat treatment was applied. Due to the high temperature processing, Cu grain growth has reached a stabilized state with a microcrystalline structure, grown from the deposited ultrafine-grained formation. After thermal processing, only limited and localized dewetting was observable (while no delamination has occured) that had, therefore, no influence on the micromechanical characterization, as the coating remained in good mechanical contact with the Si substrate.

During the mechanical data analysis, the strain rate sensitivity behavior was analyzed in detail, namely \emph{i)} yield stress as a function of strain rate and pillar diameters, and \emph{ii)} increased strain rate hardening in the HT samples at very high ($\dot{\varepsilon} = 1000$/s) strain rates. It was proposed that this behavior is a demonstration that the dislocation nucleation and motion is strongly deformation rate dependent in Cu at high strain rates.

Analysing the \emph{post mortem} surfaces of the deformed pillars it was concluded that the Al$_2$O$_3$ interlayers have successfully prevented shear localization (even in the HT pillars), resulting in smooth stress-strain curves and homogeneous deformation throughout the height of the pillars, that produced shape changes in a barrelling manner.

All these conclusions were made on FIB-milled pillars, bearing in mind that the Ga$^+$ implantation has likely impacted the resulting yield stresses. However, the proposition of such a coating to be considered as a model material to investigate the mechanical behavior of other complex (i.e. 3D printed) materials can finally be accepted by performing a large number of tests in identical conditions, and finding a good correlation between the previously reported yielding behavior and the current work.

GND densities were also studied in detail before and after compression, leading to the observation of slightly higher dislocation densities in the vicinity of the pillars' edges, resulting in an inhomogeneous distribution within the samples. This most likely plays a key role in the mechanical behavior and results in a transition of the strain rate sensitivity at HSR, when the sample sizes are reduced below a characteristic (microstructure dependent) length scale. 

Finally, no recrystallization occured in the deformed samples, regardless of their size, microstructure or ageing time, that leads to the deduction that the employed Al$_2$O$_3$ interlayers successfully prohibit such transformation even when the samples were subjected to high strain rates. This means that the proposed Cu/Al$_2$O$_3$ multi-layered system can be applied as a coating to withstand extreme conditions.

\section*{Acknowledgement}
Matthieu Lenci (Ecole des Mines, St. Etienne) is acknowledged for the TEM imaging.

\section*{CRediT statements}
\textbf{S. Kalácska:} Conceptualization, Methodology, Validation, Formal analysis, Investigation, Resources, Data Curation, Supervision, Project administration, Writing -- Original Draft, Writing -- Review \& Editing, \textbf{L. Pethő:} Investigation, Resources, Writing -- Review \& Editing, \textbf{G. Kermouche:} Resources, Writing -- Original Draft, Writing -- Review \& Editing, \textbf{J. Michler:} Resources, Writing -- Review \& Editing, \textbf{P. D. Ispánovity:} Conceptualization, Writing -- Original Draft, Writing -- Review \& Editing

\section*{Funding sources}
SK and GK were funded by the French National Research Agency (ANR) under the project No.~ANR-22-CE08-0012-01 (\emph{INSTINCT}) and No.~ANR-20-CE08-0022 (\textit{RATES}). PDI acknoledges the support by the National Research, Development and Innovation Fund of Hungary under project No.~NKFIH-FK-138975.

\section*{Data Availability}
Experimental data generated in this study have been deposited in the Zenodo database at \url{www.doi.org/10.5281/zenodo.12782449}.

\bibliography{mybibfile}

\onecolumn
\sloppy
\begin{centering}

\Large{\textbf{Supplementary Materials for} \\
Strain rate sensitivity of a Cu/Al$_2$O$_3$ multi-layered thin film}\\
\vspace{0.5cm}
\normalsize{Szilvia Kal\'{a}cska\emph{$^{a,b,*}$}, Laszlo Pethö\emph{$^{b}$}, Guillaume Kermouche\emph{$^{a}$}, Johann Michler\emph{$^{b}$}, Péter D. Ispánovity\emph{$^{c}$} }
\vspace{0.5cm}
\\
\small{\emph{$^{a}$ Mines Saint-Etienne, Univ Lyon, CNRS, UMR 5307 LGF, Centre SMS, 158 cours Fauriel 42023 Saint-Étienne, France}\\
\emph{$^{b}$ Empa, Swiss Federal Laboratories for Materials Science and Technology, Laboratory for Mechanics of Materials and Nanostructures, Feuerwerkerstrasse 39, 3602 Thun, Switzerland}\\
\emph{$^{c}$ Eötvös Loránd University, Pázmány P. stny 1/a, 1117 Budapest, Hungary}\\}
\vspace{0.5cm}

\end{centering}

\renewcommand{\figurename}{Suppl. Figure S}
\renewcommand{\tablename}{Suppl. Table T}
\setcounter{figure}{0} 
\renewcommand\thesection{SS\arabic{section}}
\setcounter{section}{0} 
\renewcommand\bibname{Suppl. References}

Data repository available at: \url{www.doi.org/10.5281/zenodo.12782449}

\begin{figure*}[!ht]
    \centering
    \includegraphics[width=0.45\textwidth]{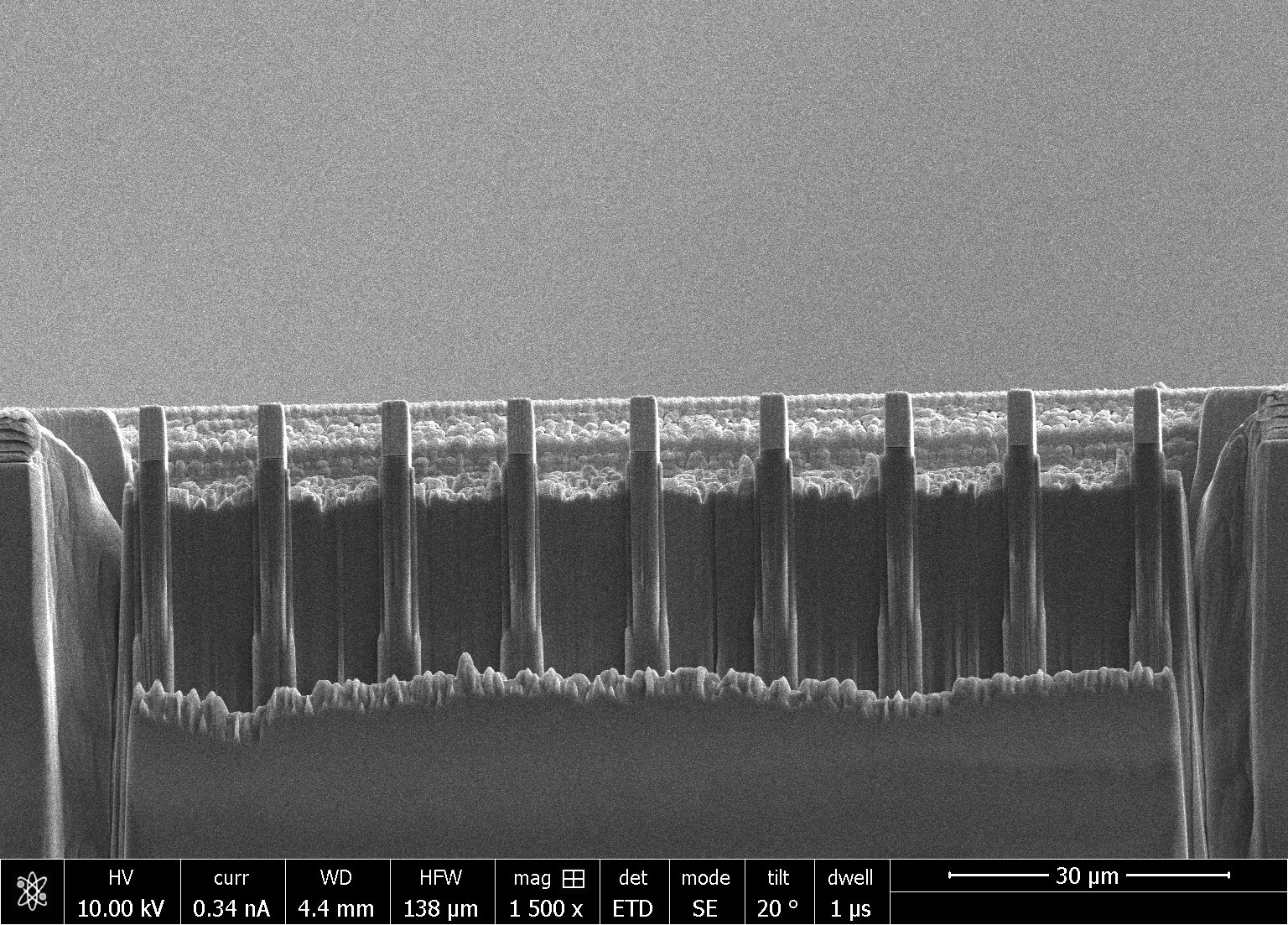}
    \includegraphics[width=0.45\textwidth]{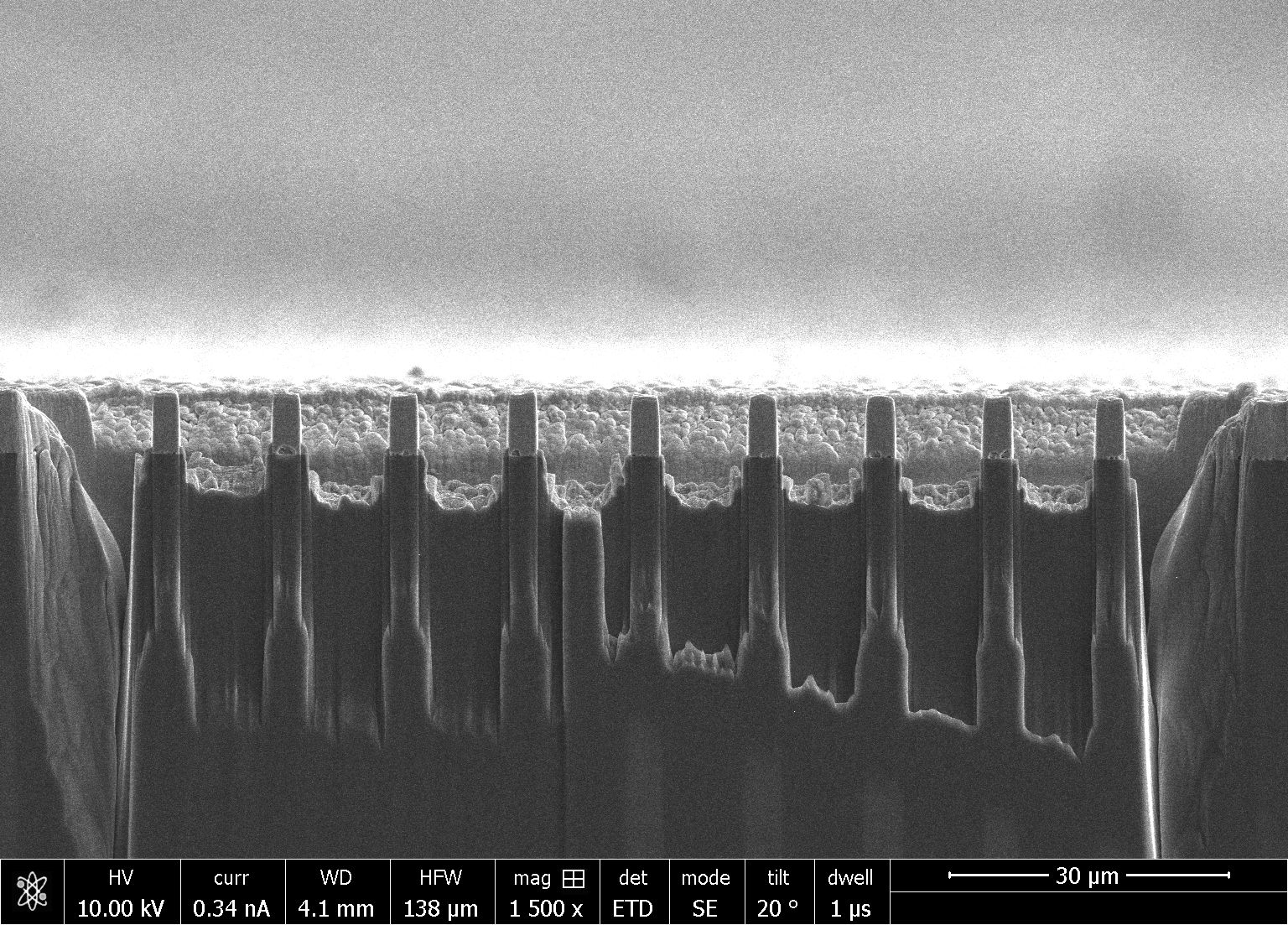}
    \caption{\textbf{overview}, 2.9 $\upmu$m (small) pillars: (left) As Deposited, (right) Heat Treated.
    \label{fig:S00}}
\end{figure*}

\begin{figure*}[!ht]
    \centering 
    \includegraphics[width=0.5\textwidth]{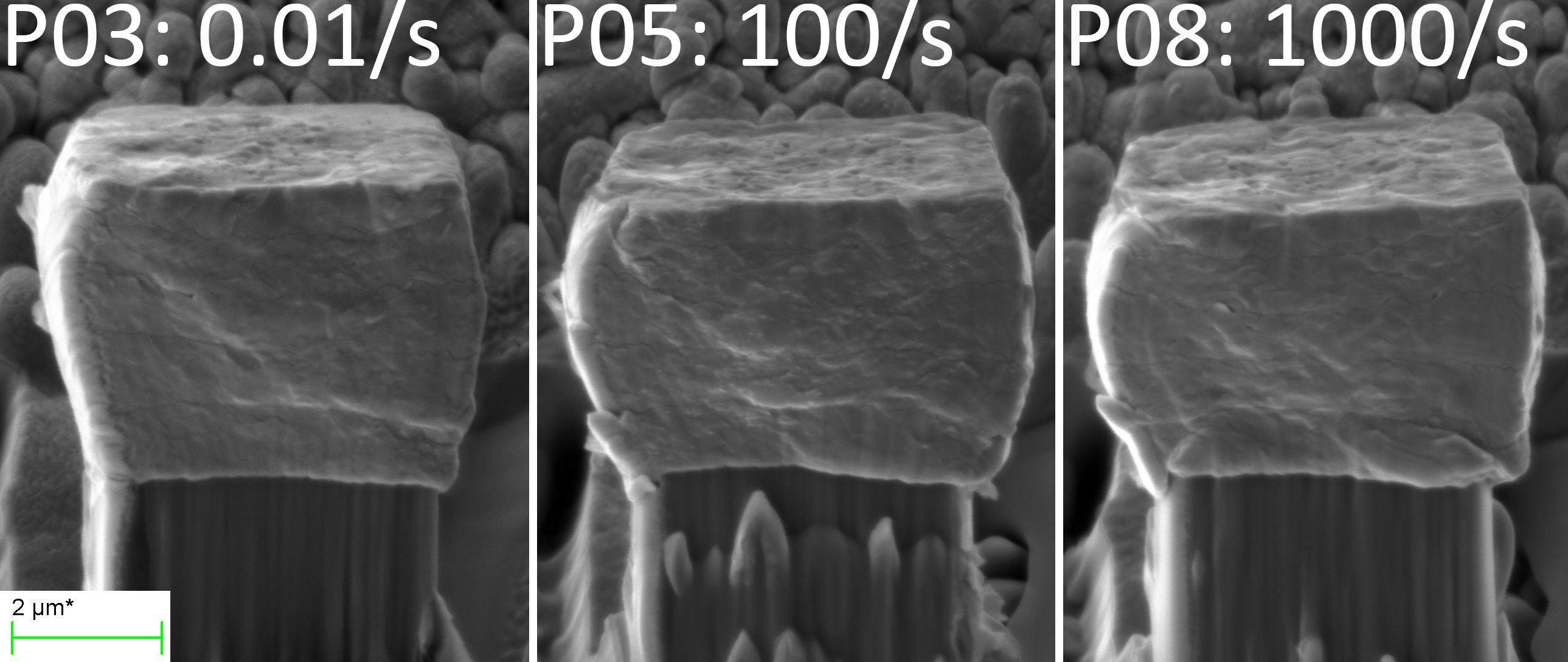}
    \caption{\textbf{As Deposited pillars} after deformation.
    \label{fig:S02}}
\end{figure*}

\begin{figure*}[!ht]
    \centering
    \includegraphics[width=0.4\textwidth]{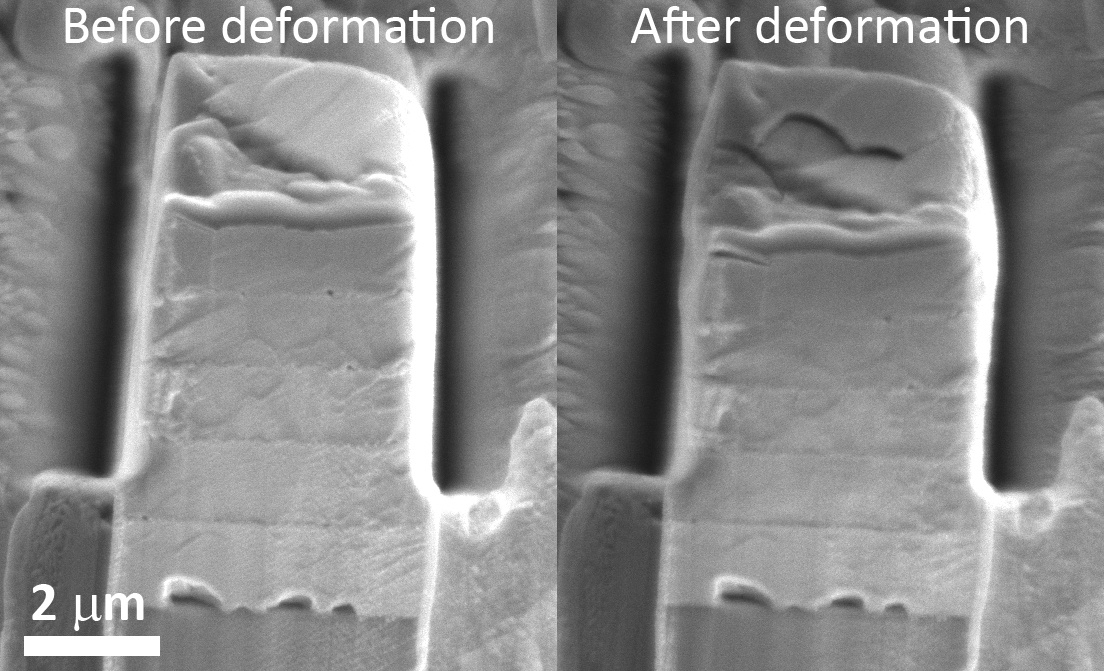}    
    \caption{\textbf{Heat Treated pillar P03,} before and after deformation.
    \label{fig:S03}}
\end{figure*}

\begin{figure*}[!ht]
    \centering
    \begin{picture}(5,0)
    \put(0,120){\sffamily{a)}}
    \end{picture}
    \includegraphics[width=0.49\textwidth]{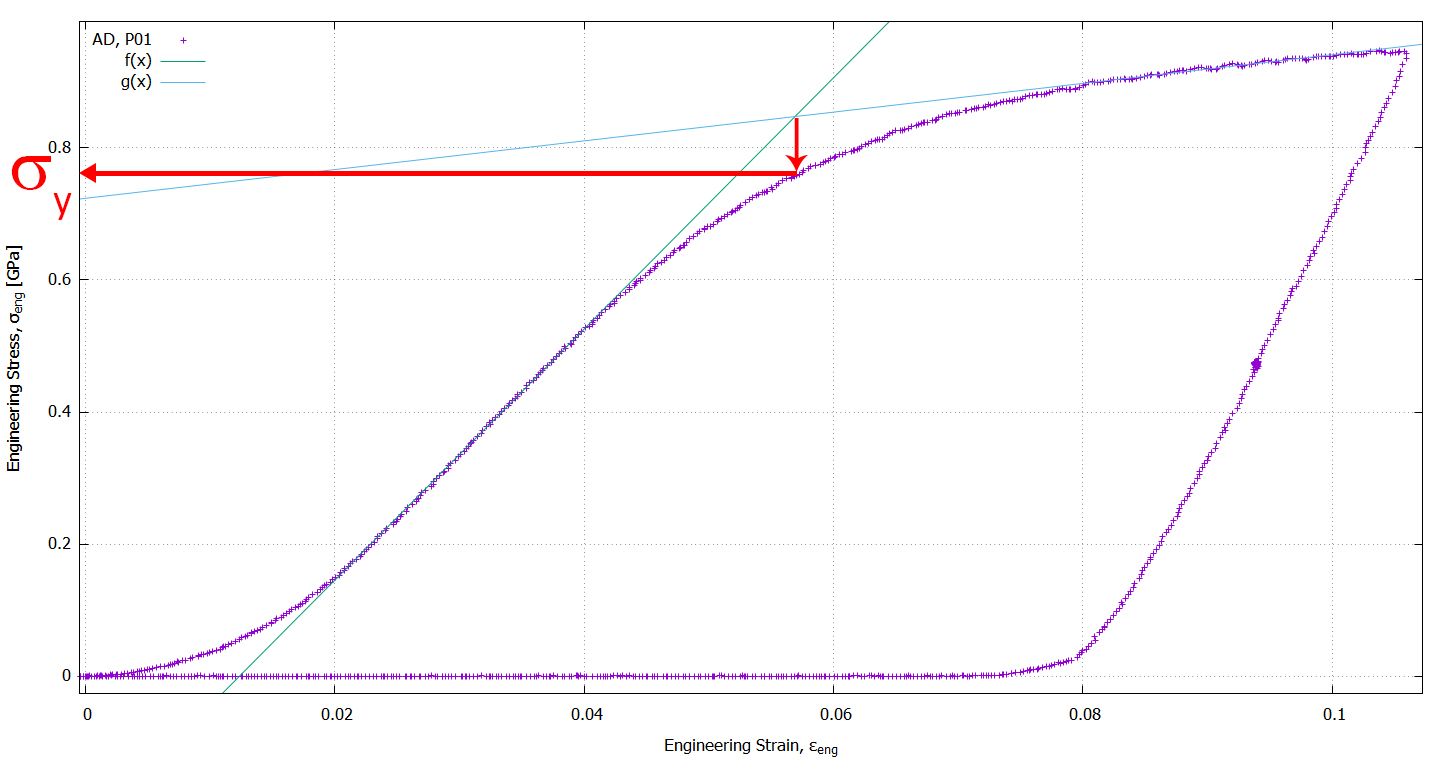}  
    \begin{picture}(5,0)
    \put(0,120){\sffamily{b)}}
    \end{picture}
    \includegraphics[width=0.45\textwidth]{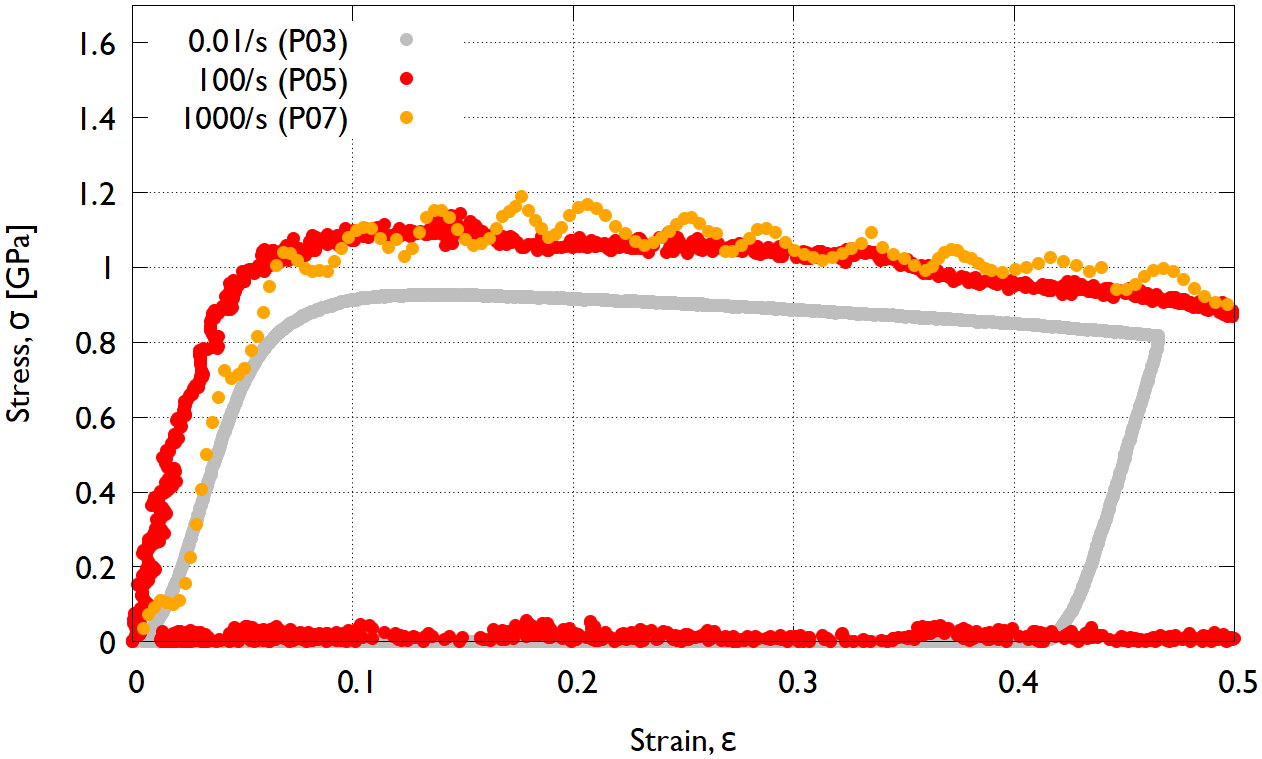} 
    \caption{\textbf{a)} Yield stress determination procedure, pillar ``AD" P01. This method was used to be consistent even at high strain rates, where the "ringing effect" introduces oscillations to the stress-strain curve. \textbf{b)} Figure \ref{fig:02}a engineering stress -- engineering strain values converted to $\sigma_{True}-\varepsilon_{True}$ curves. The applied compliance correction is $displacement-load \times C_i$, where $C_{STD}=3.695 \times 10^{-6}$, $C_{MLC}=8 \times 10^{-7}$ and $C_{ST}=9.5 \times 10^{-7}$.
    \label{fig:S01}}
\end{figure*}

\begin{figure*}[!ht]
    \centering
    \begin{picture}(5,0)
    \put(0,145){\sffamily{a)}}
    \end{picture}
    \includegraphics[width=0.45\textwidth]{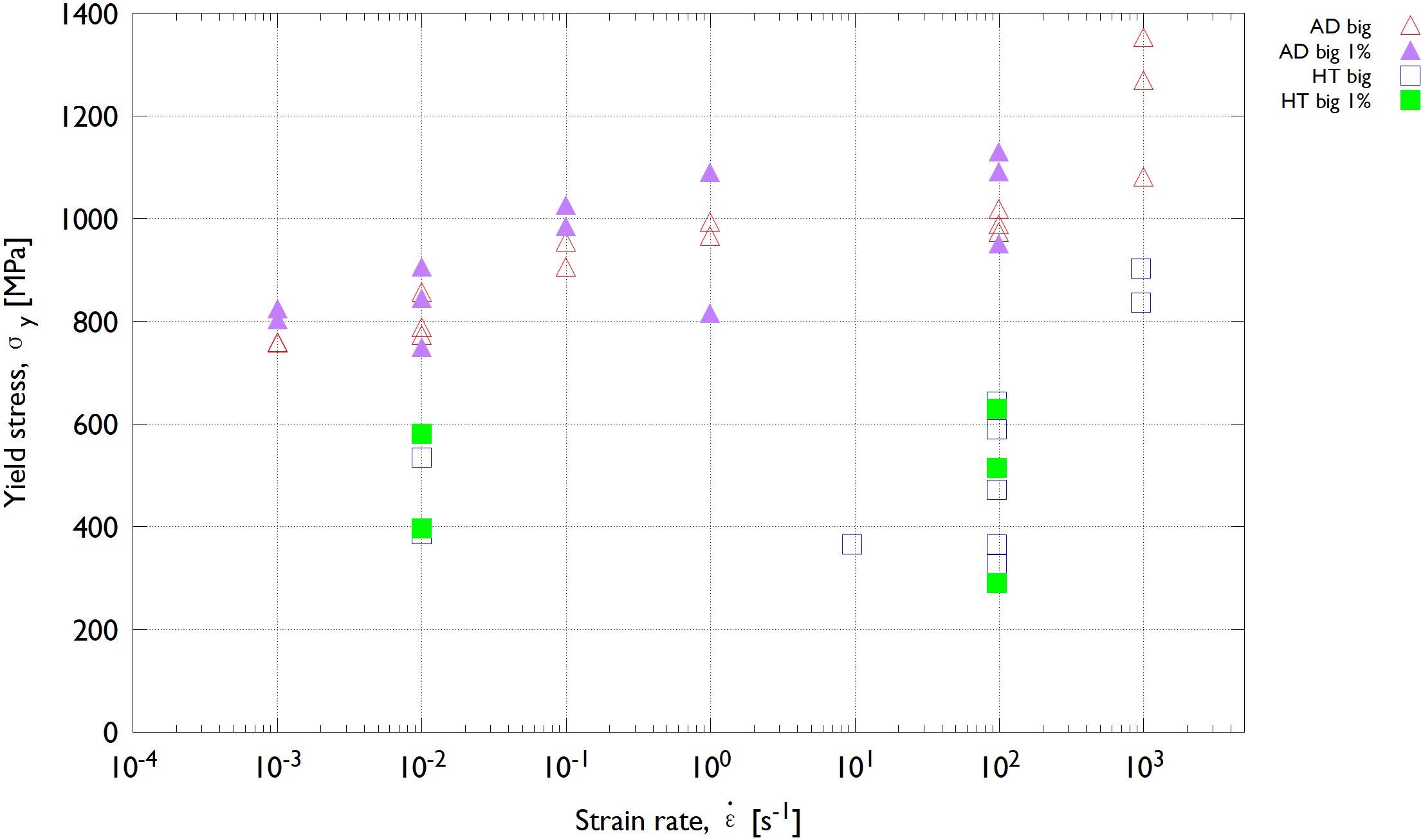}  
    \begin{picture}(5,0)
    \put(0,145){\sffamily{b)}}
    \end{picture}
    \includegraphics[width=0.45\textwidth]{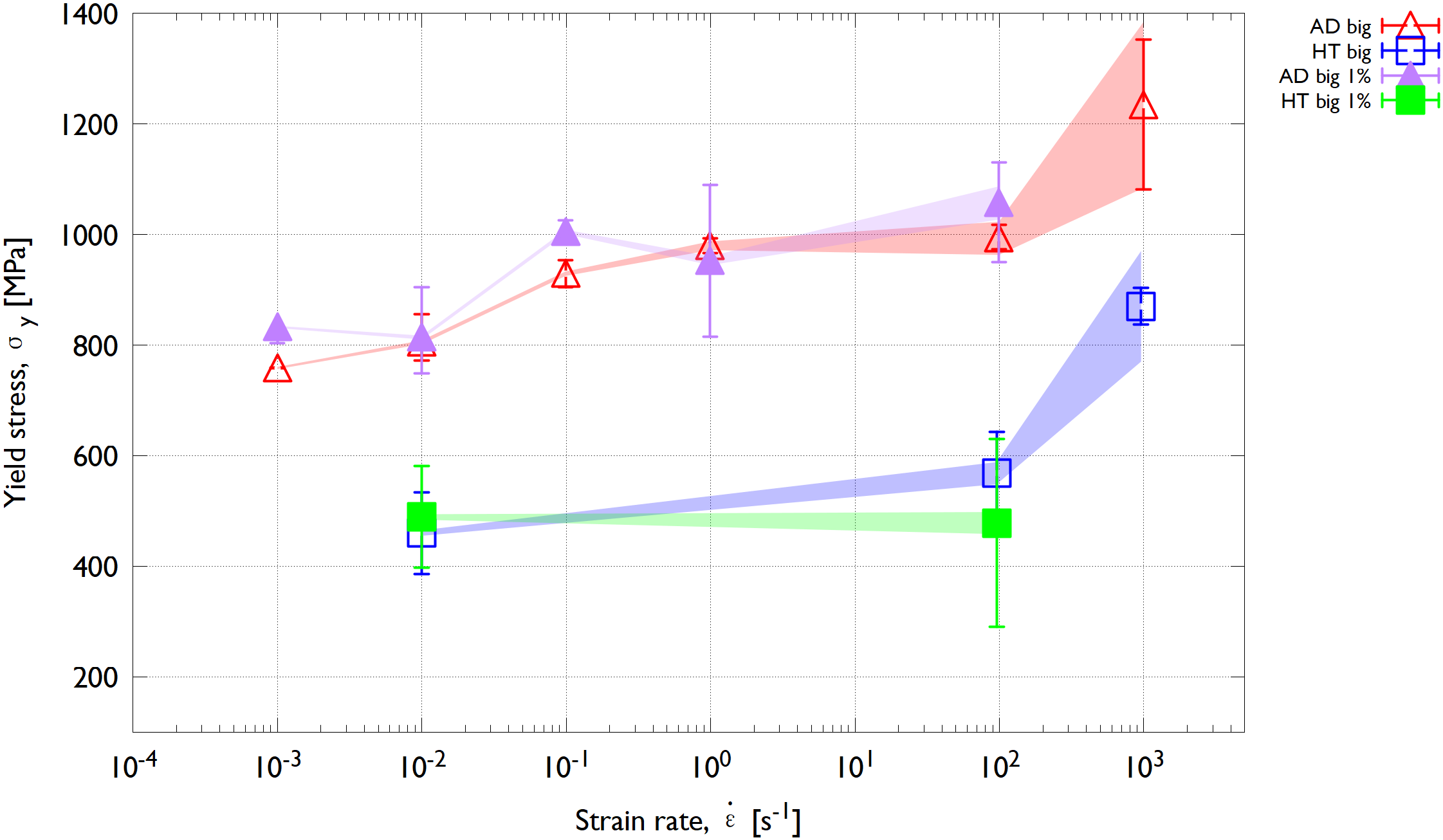} 
    \\
    \vspace{10 pt}
    \begin{picture}(5,0)
    \put(0,160){\sffamily{c)}}
    \end{picture}
    \includegraphics[width=0.55\textwidth]{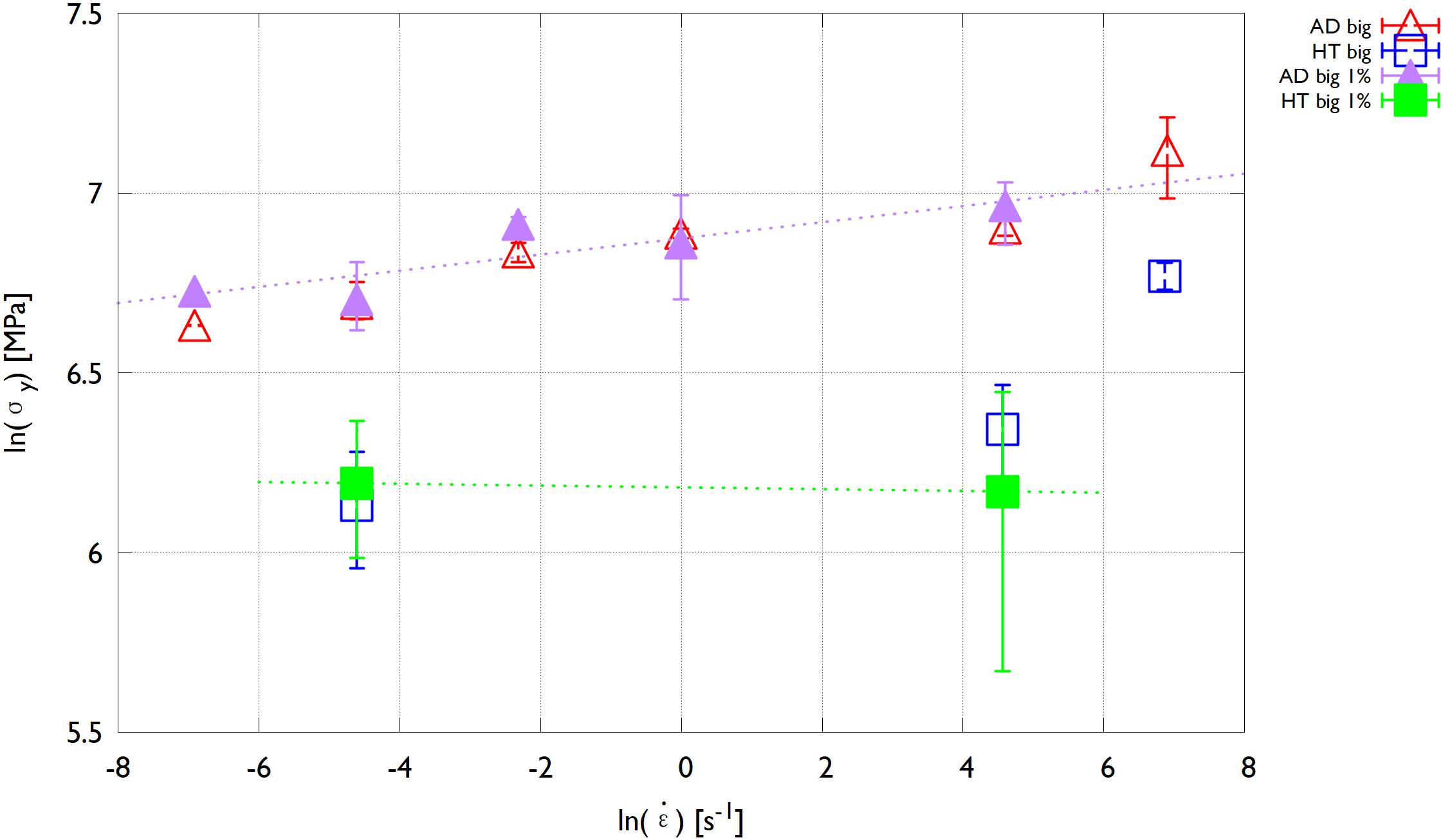} 
    \caption{\textbf{a)} Comparison of the individual yield stresses determined by the linear fitting method (empty signs, used in Fig. \ref{fig:03}) and the conventional 1\% plastic strain offset method (filled signs). \textbf{b)} (Same as a), but here the average values were plotted (same as Fig.\ref{fig:03}a), excluding the 1000/s results. \textbf{c)} ) (Same as Fig.3b). The 1\% method results in a similar m value for the AD case ($m_{\mathrm{AD}} = 0.02 \pm 0.01)$, however, the HT evaluation becomes much more noisy, resulting in a negative $m_{\mathrm{HT}}$ value ($m_{\mathrm{HT}} = -0.003)$, that is physically not expected and originating from the increased error in the determined $\sigma_y$.
    \label{fig:s08}}
\end{figure*}

\begin{figure*}[!ht]
    \centering
    \includegraphics[width=0.99\textwidth]{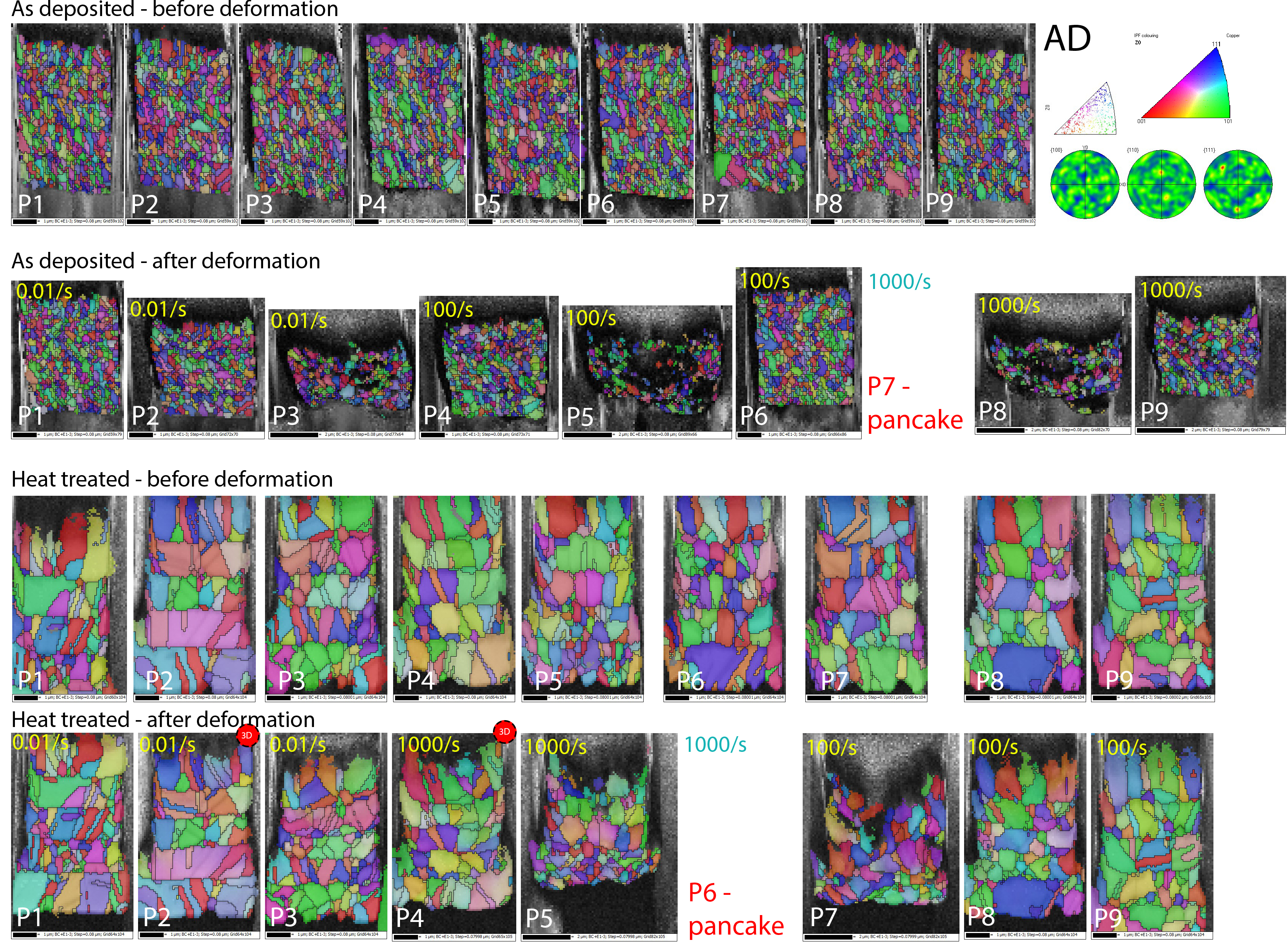}
    \caption{\textbf{EBSD results}, 4 $\upmu$m (big) pillars. Z directional inverse pole figure (IPF Z) colouring is shown together with band contract (BC, greyscale) imaging, before and after deformation. In the to right corner the full IPF triange is shown along with the registered grain colours plotted inside the IPF triangle for the as deposited (AD) case. Pole figures plotted for the X, Y and D directions show no specific texture in the AD sample. Red dots on the EBSD maps highlight the two pillars that were analysed by \textit{post mortem} 3D HR-EBSD.
    \label{fig:S04}}
\end{figure*}

\begin{figure*}[!ht]
    \centering
    \includegraphics[width=0.78\textwidth]{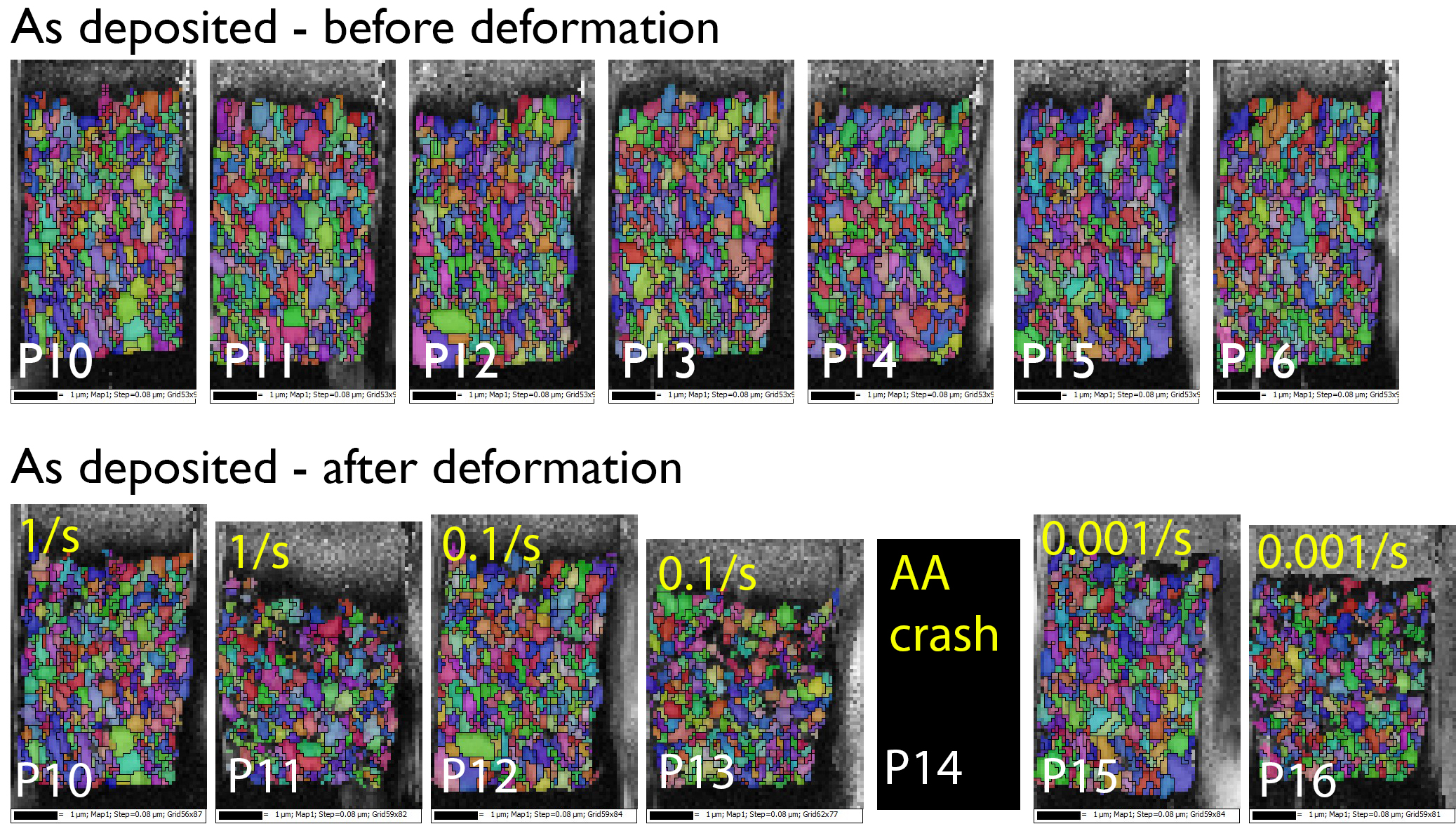}
    \caption{\textbf{EBSD results}, 4.0 $\upmu$m pillars, 2nd batch (for more statistics). Z directional inverse pole figure (IPF Z) colouring is shown together with band contract (BC, greyscale) imaging, before and after deformation. "AA crash" refers to the accidental touch of the pillar upon auto approach procedure, which resulted in an unknown deformation state, therefore this pillar was excluded from the analysis.
    \label{fig:S05}}
\end{figure*}

\begin{figure*}[!ht]
    \centering
    \includegraphics[width=0.80\textwidth]{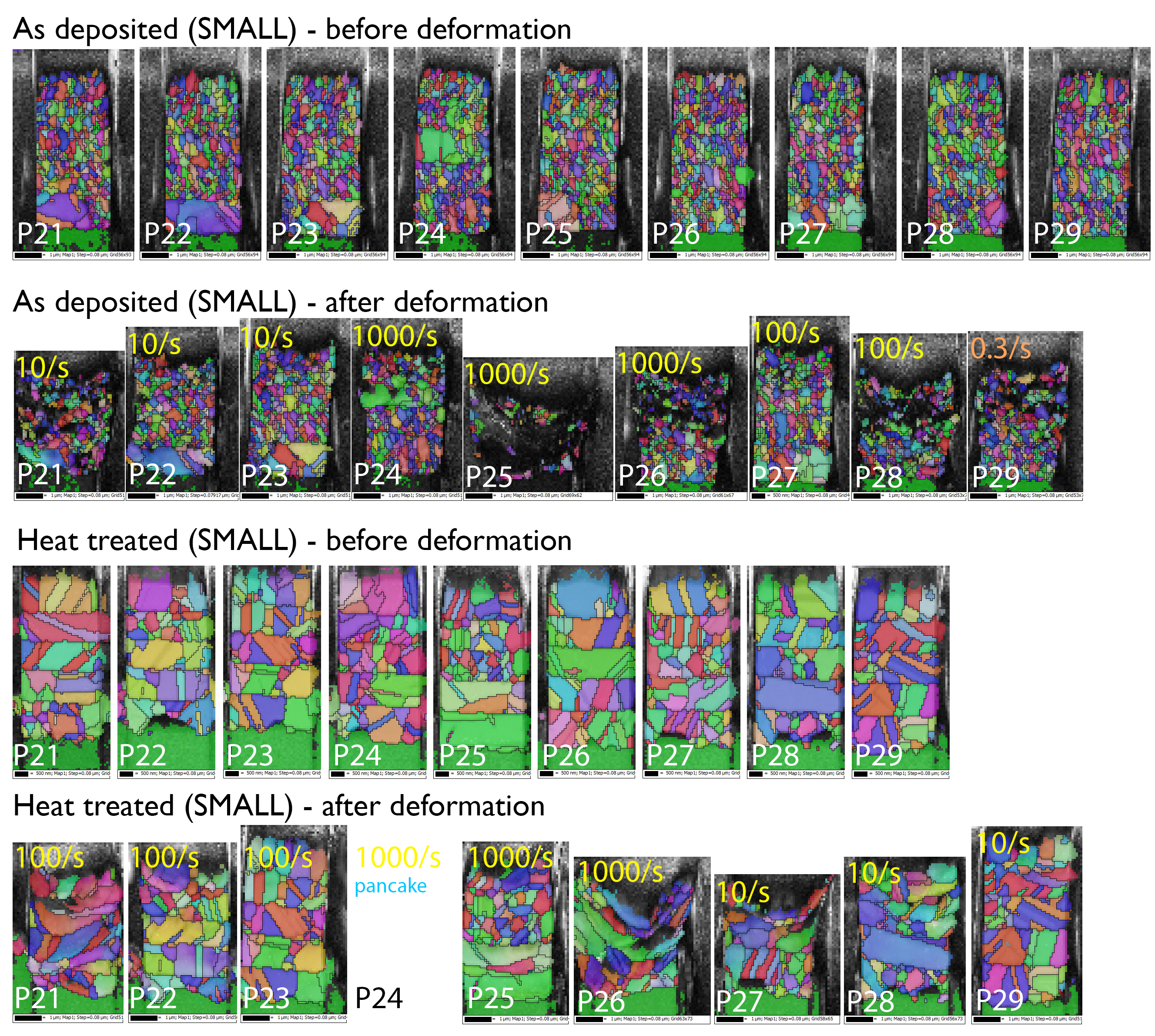}
    \caption{\textbf{EBSD results}, 2.9 $\upmu$m pillars (for size effect). Z directional inverse pole figure (IPF Z) colouring is shown together with band contract (BC, greyscale) imaging, before and after deformation.\label{fig:S06}}
\end{figure*}

\begin{figure*}[!ht]
    \centering
    \includegraphics[width=0.4\textwidth]{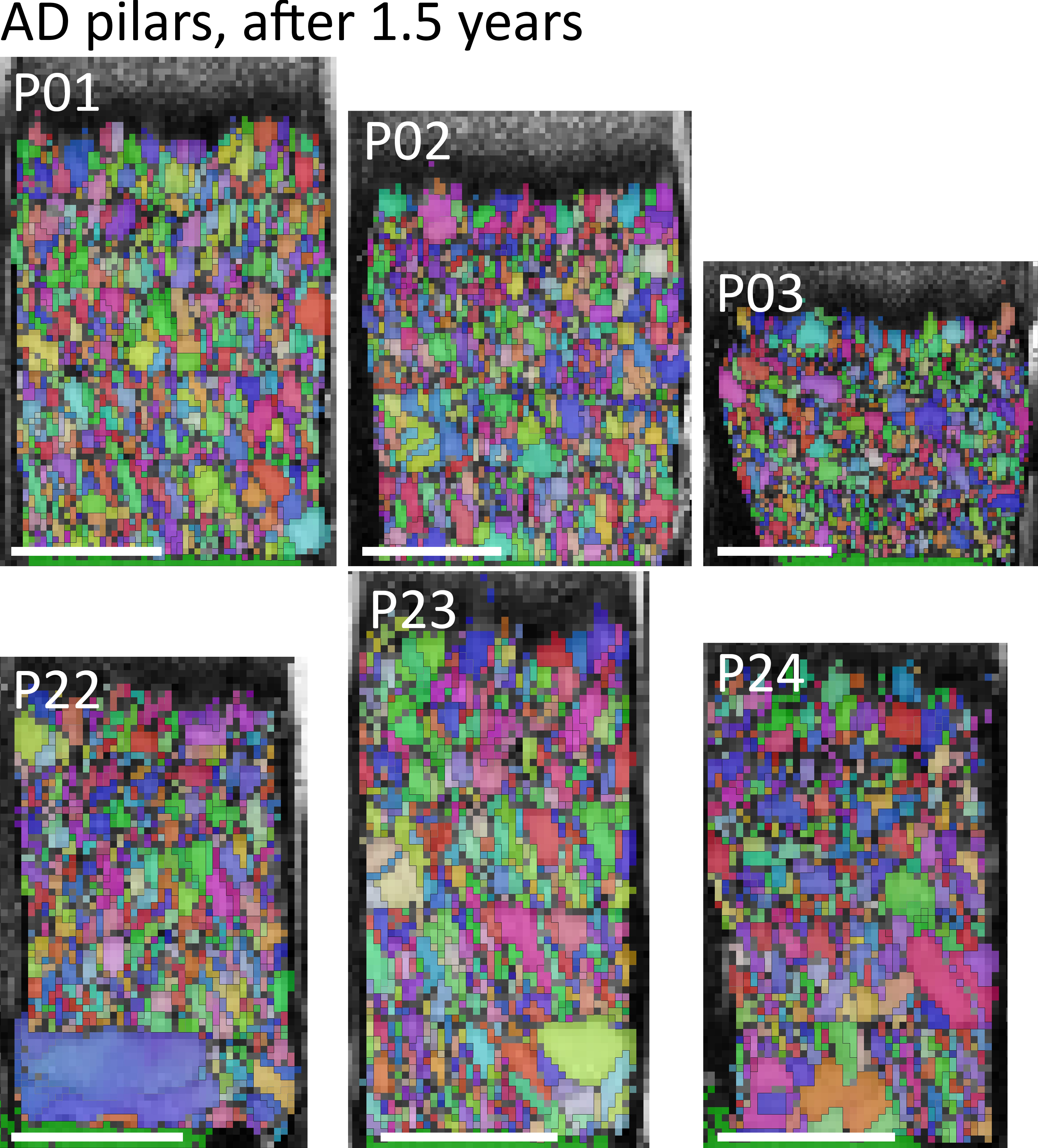}    
    
    \caption{\textbf{EBSD results of the deformed AD pillars}, aged for 1.5 years, repolished by FIB.
    \label{fig:S07}}
\end{figure*}

\begin{figure}[!ht]
    \centering
    \includegraphics[width=0.95\textwidth]{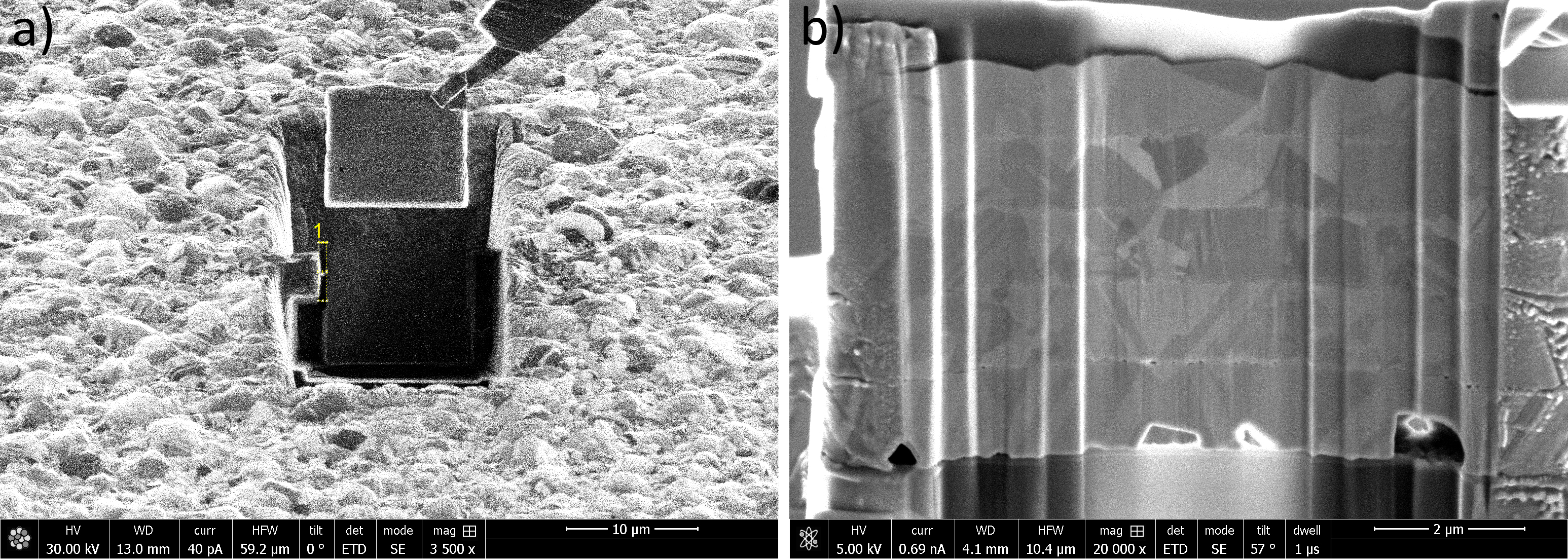}
    \caption{\textbf{TEM lamella preparation.} \textbf{a)} snapshot during the liftout process, \textbf{b)} image taken during the thinning.
    \label{fig:HTss}}
\end{figure}

\begin{figure}[!ht]
    \centering
    \includegraphics[width=0.95\textwidth]{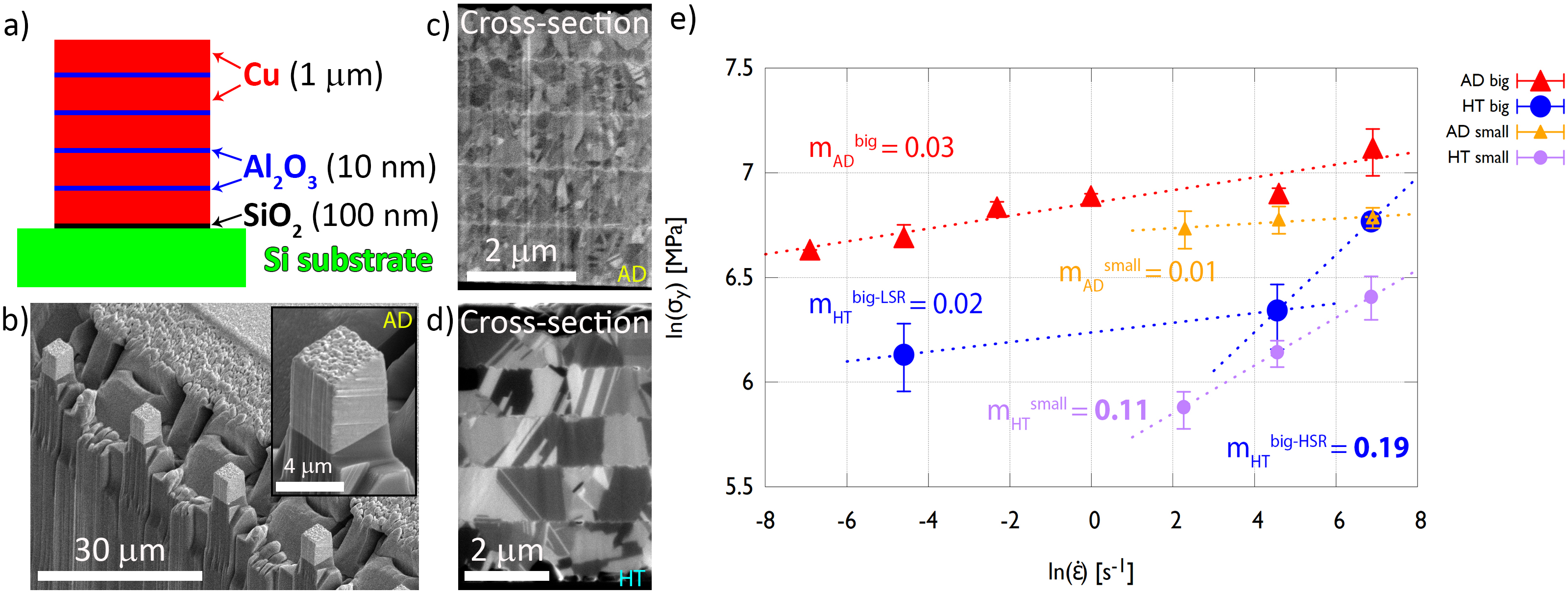}
    \caption{\textbf{Graphical Abstract}}
    \label{fig:GA}
\end{figure}

\end{document}